\PassOptionsToPackage{bookmarks={false}}{hyperref}
\documentclass[sigconf]{acmart}
\renewcommand\footnotetextcopyrightpermission[1]{} 
\usepackage{booktabs}
\usepackage{xspace}
\usepackage{enumitem}
\usepackage{listings}
\usepackage{pifont}
\usepackage{fancyvrb}
\usepackage{xcolor}

\definecolor{codered}{HTML}{D9534F}
\definecolor{codegreen}{HTML}{5CB85C}
\definecolor{codeblue}{HTML}{0000FF}

\definecolor{m1}{RGB}{224, 250, 255}
\definecolor{m2}{RGB}{219, 224, 255}
\definecolor{m3}{RGB}{252, 255, 220}
\definecolor{m4}{RGB}{249, 214, 177}

\usepackage{algorithm}
\usepackage{algorithmicx}
\usepackage{algpseudocode}
\usepackage{amsmath,amssymb}
\usepackage{color, colortbl}

\newcommand{\ie}{\textit{i.e.,} }
\newcommand{\eg}{\textit{e.g.,} }
\newcommand{\etc}{\textit{etc.} }
\newcommand{\tool}{\textsc{Hercules}\xspace}
\newcommand{\toolsh}{\textsc{Hercules-SH}\xspace}
\newcommand{\toolc}{\textsc{Hercules-FixedContext}\xspace}
\newcommand{\toolh}{\textsc{Hercules-MinusHistory}\xspace}
\newcommand{\tooli}{\textsc{Hercules-Incr}\xspace}

\newcommand{\bugsjar}{Bugs.jar\xspace}
\newcommand{\dfj}{Defects4J\xspace}

\newcommand{\elixir}{\textsc{Elixir}\xspace}

\makeatletter
\newcommand{\IS}{$\leftarrow$}
\newcommand{\func}[1]{\textsc{#1}\texttt{(}\checknextarga}
\newcommand{\checknextarga}{\@ifnextchar\bgroup{\gobblenextarga}{\texttt{)}}}
\newcommand{\checknextargb}{\@ifnextchar\bgroup{\gobblenextargb}{\texttt{)}}}
\newcommand{\gobblenextarga}[1]{#1\checknextargb}
\newcommand{\gobblenextargb}[1]{, #1\checknextargb}
\makeatother
\algnewcommand\algorithmicforeach{\textbf{for each}}
\algdef{S}[FOR]{ForEach}[1]{\algorithmicforeach\ #1\ \algorithmicdo}

\definecolor{mpcolor}{rgb}{0.1,0.6,0.1}

\definecolor{rscolor}{rgb}{0.1,0.1,0.9}

\definecolor{sscolor}{rgb}{0.9,0.45,0.1}

\definecolor{todocolor}{rgb}{1,0.3,0.3}

\newcommand{\numfixedbugs}{49\xspace}
\newcommand{\numincorrectfixedbugs}{23\xspace}
\newcommand{\numsinglebugs}{34\xspace}
\newcommand{\numincorrectsinglebugs}{21\xspace}
\newcommand{\nummultibugs}{15\xspace}
\newcommand{\numnewbugs}{13\xspace}
\newcommand{\numnewmultibugs}{9\xspace}

\begin{document}

\title{Harnessing Evolution for Multi-Hunk Program Repair}

\author{Seemanta Saha$^\dagger$}\thanks{$^\dagger$ This work was done when the author was an intern at Fujitsu Laboratories of America.}
\affiliation{
\institution{University of California, Santa Barbara}
}
\email{seemantasaha@cs.ucsb.edu}

\author{Ripon K. Saha}
\affiliation{
\institution{Fujitsu Laboratories of America, Inc.}
}
\email{rsaha@us.fujitsu.com}

\author{Mukul R. Prasad}
\affiliation{
\institution{Fujitsu Laboratories of America, Inc.}
}
\email{mukul@us.fujitsu.com}

\begin{abstract}
Despite significant advances in automatic program repair (APR) techniques over the past decade, practical deployment remains an elusive goal. One of the important challenges in this regard is the general inability of current APR techniques to produce patches that require edits in multiple locations, \ie multi-hunk patches. In this work, we present a novel APR technique that generalizes single-hunk repair techniques to include an important class of multi-hunk bugs, namely bugs that may require applying a substantially similar patch at a number of locations. We term such sets of repair locations as evolutionary siblings -- similar looking code, instantiated in similar contexts, that are expected to undergo similar changes. At the heart of our proposed method is an analysis to accurately identify a set of evolutionary siblings, for a given bug. This analysis leverages three distinct sources of information, namely the test-suite spectrum, a novel code similarity analysis, and the revision history of the project. The discovered siblings are then simultaneously repaired in a similar fashion. We instantiate this technique in a tool called \tool{} and demonstrate that it is able to correctly fix \numfixedbugs bugs in the Defects4J dataset, the highest of any individual APR technique to date. This includes \nummultibugs multi-hunk bugs and overall \numnewbugs bugs which have not been fixed by any other technique so far.
\end{abstract}



\maketitle

\section{Introduction}
\label{sec:intro}

The past decade has seen significant research activity on automatic program repair (APR) techniques~\cite{APRSurvey:2018, Monperrus:2018}. These techniques bear the promise of helping automate the otherwise laborious process of debugging and patching bugs. However, this promise is yet to be realized in terms of practical deployment of APR techniques. One reason for this is perhaps the relatively limited classes of bugs that current state-of-the-art APR techniques can correctly fix~\cite{le2013current, realbug:ICSE2015}. In particular, with one notable exception~\cite{Angelix:ICSE2016}, most APR techniques are designed to target single-hunk bugs -- bugs with patches confined to a single contiguous chunk of code, at a single location. However, the vast majority of bug patches span multiple hunks. For instance, 64\% of the bugs in the Defects4J dataset~\cite{just2014defects4j} and 76\% of bugs in the Bugs.jar dataset~\cite{saha2018bugs} require multi-hunk patches. Several previous works have acknowledged the challenges of doing multi-hunk repair~\cite{realbug:ICSE2015, Angelix:ICSE2016}. Any simple-minded expansion of the search space to explore general multi-hunk patches would clearly explode the repair search space.

Given a buggy program \emph{P}, failing at least one test in a test suite \emph{T}, an APR tool (implicitly or explicitly) searches a space \emph{S} of possible mutations to \emph{P} for one that allows the mutated program to pass all tests in \emph{T}.
	
In this work, we propose an APR technique that targets a specific but important class of multi-hunk repair problems. Our solution is broadly inspired by insights from two bodies of research. The first is research on detecting and using code clones~\cite{kim2005using, roy2008empirical, duala2007tracking, juergens2009code, CCFinder:TSE2002, deckard:ICSE2007, NiCad:ICPC2011, Meng:PLDI2011, LASE:ICSE2013, Rase:ICSE2015}. This research shows that code clones are plentiful in programs. Typically 5-25\%~\cite{roy2007survey} and as much as 50\% of a subject system can be comprised of cloned code~\cite{rieger2004insights}. Further, replication of bugs through code clones is a common phenomenon. Up to 10\% of code clones contain bugs with 55\% of bugs found in code clones being replicated bugs~\cite{islam2016bug}. The second body of work is APR techniques themselves, the vast majority of which, directly or indirectly, exploit the \textit{``plastic surgery hypothesis''} -- the ingredients for a repair can be obtained from existing code~\cite{barr2014plastic}. This can take the form of using program transformation schemas (which define the repair space) derived from a corpus of existing patches~\cite{PAR:ICSE2013, Relifix:ICSE2015, antiPatterns:FSE2016, Genesis:FSE2017, CapGen:ICSE2018, SimFix:ISSTA2018}. Alternatively, repair ingredients, such as program elements, expressions, statements, or whole snippets can be mined from existing code and re-purposed for creating a patch~\cite{GenProg:ICSE2012, ACS:ICSE2017, ssFix:ASE2017, SimFix:ISSTA2018}. These two bodies of work point to the phenomenon that it is plausible to find ``similar looking" pieces of code across a project, bearing similar kinds of bugs and warranting similar patches. Our work exploits this general insight as well.
 	
Our proposed method generalizes single-hunk repair techniques to include bugs that may require applying a substantially similar patch at a number of locations, \ie a multi-hunk patch. We term the underlying set of repair locations as \emph{evolutionary~\footnote{We use evolution as a metaphor for the environment, \ie context, of a piece of code, in addition to the changes it undergoes over its lifetime.} siblings} -- similar looking code, instantiated in similar contexts, that are expected to undergo similar changes, over the lifetime of the codebase. It is important to note, as established in RQ1 (Section~\ref{repair-localization}) that these evolutionary siblings are not simply code clones, in the traditional sense. Further, our approach is orthogonal to use of the plastic surgery hypothesis which involves mining existing code (donor) for abstract schemas or concrete ingredients from which to compose the present repair (the donee), \ie a donor-donee relationship. By contrast, our proposed analysis seeks to find evolutionary siblings exposed by the current bug, which can be repaired simultaneously and in a similar fashion. This is, in principle, independent of the actual technique used to perform the repair, and quite compatible with a donor-donee repair mechanism.

The key to our approach is to be able to accurately identify \emph{evolutionary siblings} for a given bug. This presents the following technical challenges:

\textbf{\textit{Challenge 1:}} Evolutionary siblings are not simply code clones. Thus, further analysis is required to expose the desired sibling relationships. 

\textbf{\textit{Challenge 2:}} The spectrum generated by the test cases may not expose or even cover all the sibling instances. Missing some of the siblings can produce a partial repair at best.

\textbf{\textit{Challenge 3:}} Any imprecision in identifying these siblings can be potentially fatal to identifying a successful repair. Again, an under-approximation can produce a partial repair at best. Also, it is simply not computationally feasible to search the power set of an over-approximate set of potential siblings. 

At the heart of our proposed method is an analysis that uses three distinct sources of information to accurately identify evolutionary siblings suitable for repair, for the bug at hand. First, it uses the test spectrum to implicate one or more of the siblings to provide a starting reference for the sibling identification. Second, it identifies all siblings of the reference sibling that have syntactic similarity but \emph{also} semantic similarity of their context. This is done using a code similarity analysis that combines syntactic similarity with a limited scope data flow analysis to enforce similarity of the \emph{semantic context} for identified siblings. This code similarity analysis may, in principle, identify potential siblings that are outside the scope of the test-spectra. This is an essential feature of our analysis that compensates for the weakness and incompleteness of typical test suites. Developers may often not add test-cases witnessing the bug for \emph{each} of the siblings but rather only for some, or only one of them. Third, our method uses the revision history information to further enforce that the siblings thus identified bear a similar history of changes. This third feature discards false positives, that are not necessarily siblings in a co-evolutionary sense. Once evolutionary siblings are identified they can be handed off to any repair algorithm that should enforce \emph{simultaneously} generating a \emph{substantially similar repair} (modulo namespace variations) for all siblings. In principle, any traditional repair tool could be suitably retro-fitted to perform this part.

We have implemented the proposed technique in a tool \tool{}~\footnote{Our tool kills multi-location bugs like the mythical Hercules killed the multi-headed monster Hydra.} 
and evaluated it on the widely used Defects4J dataset. In our experiments \tool{} was able to correctly fix \numfixedbugs bugs, the highest of any single APR technique so far. This includes \nummultibugs multi-hunk bugs, and overall \numnewbugs bugs which have not been fixed by any other technique so far.

It is noteworthy that, although not specifically discussed in \cite{ACS:ICSE2017} or \cite{SimFix:ISSTA2018}, the implementations of ACS and SimFix are in fact capable of performing simple instances of multi-hunk repair. The bug repair counts reported in \cite{ACS:ICSE2017} and \cite{SimFix:ISSTA2018} include such bugs. Specifically, these tools perform multi-hunk repairs when there are separate test-cases separately implicating each bug location. Then the tool simply iteratively repairs these bugs independently, one after another. This potentially increases the search space exponentially in the number of locations, but still turns out to be viable for the simplest instances. By contrast, our approach is much more general, exploiting the sibling relationship to keep the search space effectively the same as a single-location patch. It also patches locations not covered by the test spectrum. Angelix~\cite{Angelix:ICSE2016} also performs multi-hunk repairs but again it does not exploit any relationship between the patched locations, beyond any dependencies inadvertently captured by the test suite. 
The main contributions of this paper are:
\begin{itemize}
	\item \textbf{\textit{Technique:}} An APR technique generalizing traditional single-hunk repair to bugs that may require applying a substantially similar patch at a number of locations
	\item \textbf{\textit{Analysis:}} An analysis implementing accurate detection of evolutionary siblings in service of the above repair goal
	\item \textbf{\textit{Tool:}} An instantiation of the proposed repair technique in a tool \tool
	\item \textbf{\textit{Evaluation:}} An evaluation of \tool{} on the Defects4J dataset
\end{itemize}

The rest of the paper is organized as follows. Section~\ref{sec:background} presents basic background material to orient the reader, followed by a motivating example illustrating our approach in Section~\ref{sec:motivating_example}. Section~\ref{sec:approach} presents our proposed approach in detail. Sections~\ref{sec:exp_setup} and \ref{sec:results} present our experimental set-up and evaluation respectively. Section~\ref{sec:discussion} discusses the limitations of our approach. followed by a discussion of related work in Section~\ref{sec:related_work}. Section~\ref{sec:conclusion} concludes the paper. 

\textbf{\textit{Relationship to the \tool{} ICSE2019 paper~\cite{Hercules:ICSE2019}:}} This manuscript reports on the latest experimental evaluation of \tool{} as originally published~\cite{Hercules:ICSE2019}. Specifically, while the technique is essentially unchanged from \cite{Hercules:ICSE2019}, some bug-fixes in the implementation led to \tool{} generating some new patches, both correct and incorrect, which is reflected in the results reported in Section~\ref{sec:results}. Further, Tables~\ref{tbl:results_by_bug_id_sh} and \ref{tbl:results_by_bug_id_mh} also provide the Bug IDs of all correct and plausible patches generated by \tool{} as well as those of several recent APR techniques, for which results are available.

\section{Terminology}
\label{sec:background}

\subsection{Terminology \& Definition}
In this paper, we consistently use the following terminology and definitions.

\textbf{Single vs. multi-hunk bugs.} Single-hunk bugs require program edits (insertions, deletions, or modifications) at a single location or a set of contiguous locations. Multi-hunk bugs require program edits at multiple non-contiguous locations.

\textbf{Repair location.} A program statement that we want to modify, delete, or insert a new statement. It should be noted a repair location may or may not be the actual buggy location.

\textbf{Repair schema.} An abstract program transformation template, such as adding a null checker or inserting a method invocation to a given repair location.

\textbf{Candidate patch.} A concrete modification to a program realized by instantiating a repair schema. 

\textbf{Repair space.} The pool of generated candidate patches. The size of the repair space refers to the number of generated candidate patches.

\textbf{Plausible patch.} A plausible patch is one that simply passes all test-cases in the test suite. It should be noted that a plausible patch may still be incorrect because the test-suite may provide an incomplete specification.

\textbf{Correct vs. incorrect patch.} We classify a patch as \textit{correct}, if it is semantically equivalent to the developer-provided patch, based on a manual examination. This is consistent with the definition used in previous work~\cite{Kali:ISSTA2015, SPR:FSE2015, Durieux:CoRR2015, Prophet:POPL2016, Angelix:ICSE2016}. An \textit{incorrect patch} is a patch that is not correct.

\subsection{Generate and Validate Repair Approach}

Search-based repair approaches or so-called \textit{generate and validate (G\&V)} approaches, start with a buggy version of the program, a test-suite with at least one failing test case (revealing the bug) and one passing test case, and a set of repair schemas or program mutations to use to repair the program. A typical G\&V technique operates using the following basic steps:

\textbf{Step 1: Fault localization.} This step produces a ranked list of repair locations. This step is typically realized using spectrum based fault localization (SBFL) techniques such as Tarantula~\cite{Tarantula:ICSE2002}, Zoltar~\cite{Zoltar:SINTER2009}, and Ochiai~\cite{Abreu:2009}.

\textbf{Step 2: Generate candidate patches.} The repair approach examines each repair location in the fault localization list and applies each of the repair schemas on this statement, one at a time, in some order, to produce potential \textit{candidate patches}. The order of repair schemas may be decided using genetic algorithms~\cite{GenProg:ICSE2012}, random choice~\cite{RSRepair:ICSE2014}, heuristically~\cite{AE:ASE2013, SPR:FSE2015}, or using a machine-learned model~\cite{Prophet:POPL2016}. 

\textbf{Step 3: Selection of candidate patches and validation.} Each candidate patch is evaluated against the test suite, and if it passes, is output as a plausible patch. This step, which is computationally expensive, is generally optimized by first ranking candidate patches and selecting only a few of them for validation. The validation step is further optimized by first testing a candidate patch against a subset of the suite, \eg only failing tests, before executing the complete suite.

\section{Motivating Example}
\label{sec:motivating_example}

\begin{figure*}[t]
\begin{lstlisting}[frame=single,language=Java, basicstyle=\fontsize{8.4pt}{8.4pt}\selectfont\ttfamily, numberstyle=\scriptsize]
univariate/BrentOptimizer.java
public class BrentOptimizer extends BaseAbstractUnivariateOptimizer {
	:
142	|\colorbox{m2}{UnivariatePointValuePair current = new UnivariatePointValuePair(x, isMinim ? fx : -fx);}|
	:
226	      |\colorbox{m4}{current = new UnivariatePointValuePair(u, isMinim ? fu : -fu);}|
	:
                 if (checker != null) {
                     if (checker.converged(iter, previous, current)) {
230	-                   |\colorbox{m4}{\textcolor{red}{return current;}}|
230	+                   |\colorbox{m4}{\textcolor{blue}{return best(current, previous, isMinim);}}|
                     }
                 }
	:
                     }
                 }
             } else { // Default termination (Brent's criterion).
267	-               |\colorbox{m2}{\textcolor{red}{return current;}}|
267	+               |\colorbox{m2}{\textcolor{blue}{return best(current, previous, isMinim);}}|
            }
             ++iter;
         }
\end{lstlisting}
\caption{The fix for Math \dfj ID: 24 with program context}
\label{fig:math-24}
\vspace{-3mm}
\end{figure*}

In this section, we provide a brief overview of our approach with a motivating example, presented in Figure~\ref{fig:math-24}. The example is a real-world bug-fix in Apache Commons Math (Jira Official Bug ID: MATH-855), which is also a bug instance (Math-24) in the popular \dfj dataset. The bug report says that Brent Optimizer was not always reporting the best point. As we can see in Figure~\ref{fig:math-24}, the assigned developer made two similar modifications at two locations to fix the bug, and the modifications are not trivial. Previously, the method was returning an object instance $current$ from two locations but actually they should be a method call $best(current, previous, isMinim)$ at each location. Now we discuss how \tool{} fixes this bug while overcoming the research challenges outlined in Section~\ref{sec:intro}.

{\bf The repair locations may not be part of code clones.} From the developer's patch, we can easily observe that although the statements at the repair locations are the same, they are not code clones in a traditional sense. This is because the respective statements around the repair locations do not match. Generally, clone detection tools try to get a good trade-off between the number of minimum statements/tokens in a code snippet to be a clone to avoid producing a lot of false positives. Certainly, we can detect clones at a statement level. However, in that case, we would get many false positives. For our example, if we search the entire code-base using {\tt return current}, we will get 10 instances. Furthermore, if we abstract the identifier, which is indeed required for multi-hunk bug-fix since the identifier names may vary in two snippets of code, and search {\tt return \$x}, we will get many more similar statements. Therefore, it is evident that the statement-level similarity does not work here. On the other hand, if we use any traditional clone detection that uses a sliding window approach to detect clones using adjacent context, these repair locations will not be part of any detected clone snippets. 

In order to overcome the aforementioned problem, \tool{} uses the notion of semantic context instead of the syntactic neighborhood. Specifically, \tool{} uses {\it reaching definition} analysis (Section~\ref{subsec:context}) to extract statements, within the method boundary, on which the statement at the repair location has a data-flow dependence. The extracted statements represent the semantic context of the repair location statement. This mechanism also allows \tool{} to associate a variable-sized context with a repair location. Then \tool{} performs a deeper statement-level AST analysis to determine (syntactic) similarity between the set of repair locations combined with their respective semantic contexts. For example, Figure~\ref{fig:math-24} presents two contexts for two repair locations, highlighted by blue and orange color. For both hunks, the context statements are physically far from the corresponding repair location. Even more interestingly, they are interleaved. For the first hunk, the repair location is line number 230 but its reaching definition is at line number 226. For the second hunk, its reaching definition (line number 142) is 125 lines away from the repair location (line number 267). However, after \tool{} extracted the semantic context, the two resulting code snippets become similar, although not identical (declaration vs. assignment). However, \tool{} concludes that the both return statements are used in similar context, based on the AST analysis.

{\bf Weak Specification and Spurious Repair Locations.} It is well known that test suites typically do not cover every program location. This issue has specific ramifications for multi-location bug fixing. For instance, in the current example, the failing test case covers the second location but not the first location. Therefore, any repair tool that solely relies on test cases for identifying repair locations, cannot generate the complete correct patch. On the other hand, if we apply program transformations in all the similar locations identified from the previous step, it may not be appropriate either. In order to find {\it true} evolutionary siblings, \tool{} extracts the revision history of the target repair locations and analyzes whether these lines were revised independent of one another. For our example, although the first repair location is not covered by any test case, its revision history shows it was never modified independent of the second location. This allows \tool{} to confidently apply similar changes to both locations.

{\bf Rich Repair Space.} Another general limitation of G\&V approaches is dealing with an enormous number of candidate patches. The number of candidate patches increases exponentially with the number of repair locations. For example,  there are more than 14,000 repair expressions that are valid in each repair location. These repair expressions can be plugged into the program transformation schemas, resulting in thousands of candidate patches. Even if we are fortunate enough to determine the correct program transformation and API call (which itself is very difficult), there can be 18 valid concrete invocations of {\it method call} for different combinations of parameters. For two locations, the number of candidate patches would be $18 \times 18 = 324$. These statistics illustrate the enormity of the repair space for multi-hunk bug fixes, and show why a naive approach would not work in such scenarios. In order to overcome this problem, \tool{} employs a strategy of simultaneous repair of evolutionary siblings. Furthermore, inspired by other existing approaches such as Prophet~\cite{Prophet:POPL2016} and \elixir{}~\cite{Elixir:ASE2017}, \tool{} uses machine learning techniques to rank and prune most of the candidate patches.

\section{\tool}
\label{sec:approach}

\begin{figure*}
\begin{center}
\includegraphics[width=0.9\textwidth]{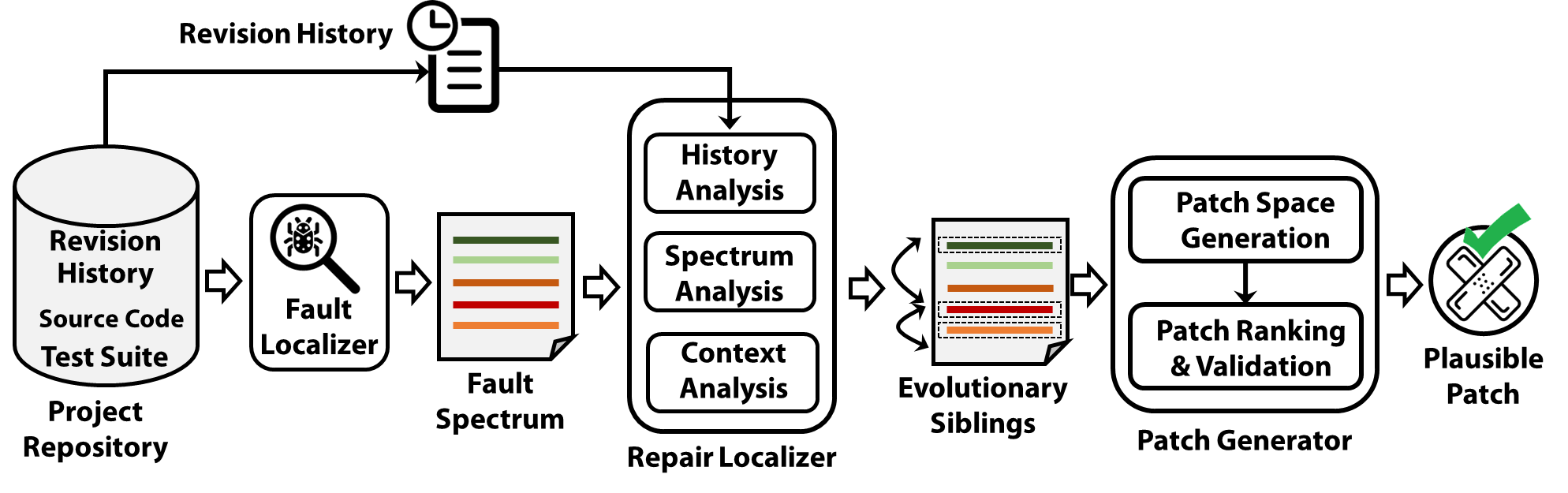}
\end{center}
\caption{Overview of \tool.}
\label{fig:elixirm}
\end{figure*}

\subsection{An Overview}
Figure~\ref{fig:elixirm} presents the basic workflow of \tool. Given that there is a bug in a program ($P$), \tool takes the source code of $P$ with its version history, a test suite ($T$) with at least a failing test case, and optionally a bug report, and generates a correct single-hunk or a multi-hunk patch that fixes $P$, in a successful run. 

\tool{} works in four major steps to mutate $P$ and eventually to generate a patch.  In the first step, \tool uses a spectrum based fault localization (SBFL) technique to identify potential repair locations. For a given repair location, in the second step, \tool{} identifies evolutionary siblings (Definition~\ref{def:siblings}) by leveraging reaching-definition (Section~\ref{subsec:context}) and version history analysis (Section~\ref{subsec:history}). If such evolutionary siblings are found, this step also produces the mapping between similar variables and objects among the evolutionary siblings. It is worth noting that \tool{} can also repair one-hunk bugs. Therefore, if a evolutionary sibling is not found for a particular repair location, the rest of the steps are continued with only one repair location. In the third step, \tool{} abstracts all the mapped variables and objects in all the repair locations, and instantiates repair schema simultaneously. After the patches are applied, the abstract variables are reverted back to their original variables. In the final step, \tool{} selects the Top $n$ candidate patches for validation. \tool{} repeats these steps for each repair location by SBFL until a plausible patch is generated or timed out. In summary, \tool{} adds a novel step in the repair process that is not available in any conventional G\&V approaches (Section~\ref{sec:background}). We refer this step as  {\it repair localization}, i.e., identifying the evolutionary siblings (if they exist) that require changes together to fix the bug.

\subsection{Identification of Evolutionary Siblings}
\label{subsec:siblings}

In this work, by {\it evolutionary siblings}, we mean the repair locations that are similar,  have used in a similar context, have similar evolution history, and at least one of the repair locations has been exercised by the fault reproducing test cases. Our motivating example in Section~\ref{sec:motivating_example} already showed that a traditional clone detector is not sufficient to identify evolutionary siblings. The fact is further supported by our results in Section~\ref{repair-localization}. \tool{} works in three steps to find evolutionary siblings. Since code matching, especially at the AST level, is expensive, \tool{} lazily applies the tree similarity algorithm first at the repair location level. Then only for similar repair locations, \tool{} extracts the relevant context, and again applies the tree similarity algorithm for context. Finally, \tool{} leverages version history to find the evolutionary siblings with high confidence. In the subsequent sections, we concretely define evolutionary siblings and describe each aforementioned step in more detail.

\subsubsection{Preliminaries} \tool{} represents and manipulates programs as abstract syntax trees $(AST)$ to analyze programs and to instantiate repair schemas. Although the $AST$ representation is standard, its relevant terms and assumptions vary based on usage, and thus deserve a concrete definition in our context. In this section, we formally define all the important terms in our context.

\begin{definition}
\label{ast} 
An {\it abstract syntax tree (AST)} is an ordered tree ($T$) where each node ($N$) represents a program element (a method or a statement). $T$ is a tuple of $<G, X, r, M>$ where $G$ is a context free grammer, $X$ is a finite set of nodes in the tree, $r$ is the root node, and $M$  maps $r$ to its each children. 
\end{definition} 

We also assume that for each node, there exists the following methods: $type(N)$ returns the type of a repair expression such as int or double, $kind(N)$ returns the kind of AST node such as statement or method invocation, $parent(N)$ returns the parent node of $N$, and $children(N)$ returns that children of $N$.
It should be noted that the number of children nodes in $N$ vary based on $kind(N)$, For example, a {\it Binary Expression} always has two children, whereas a block may have an arbitrary number of statements. Furthermore, we can also assume that all the subtrees under $T$ are also ASTs for simplicity, e.g., an AST at statement level. However, in this case, $parent(r) \neq null$.

\begin{definition}
\label{def:spectra}
\sloppy{\it Program spectra ($S$)} is a set of AST nodes $\{N_{s_1}, N_{s_2},..,N_{s_n}\}$ such that $\forall N_{s_i} \in S: kind(N_{s_i}) = statement$ and exercised by the failing test cases.
\end{definition}

\begin{definition}
\label{def:rl}
A {\it repair location ($R_l$)} is an AST node, $N_s$, where $kind(N_s)=statement$ at a location $l$.
\end{definition} 

\begin{definition}
\label{def:context}
{\it Context of relevance ($C_{R_l}$)} is a set of statements $\{N_{s_1}, {N_{s_2},.., {N_{s_p}, R_l}}\}$ with respect to a repair location ($R_l$) such that $\forall N_{s_i} \in C_{R_1} : \Pi(N_{s_i}, R_l) = true$ where $\Pi$ is a reaching-definition function.
\end{definition}

\begin{definition}
\label{def:history}
{\it Edit history of a repair location ($H_{R_l}$)} is a sequence of AST edit operations, also known as program differences, $\{D_{c_i}, D_{c_j},..D_{c_k}\}$, where $D_{c_i} = \Delta(R_{l,c_i}, R_{l,{c_{i-1}}})$ and $c_i$ denotes the $i^{th}$ commit in version history.
\end{definition} 

\begin{definition}
\label{def:siblings}
{\it Evolutionary siblings ($E_{R_l}$)} with respect to a repair location $R_l$ are a set of repair locations $\{R_l, R_{l_1}, R_{l_2},..,R_{l_n}\}$ such that $R_l \in S$ and $\forall{(R_l, R_{l_i}) \in E : \zeta(C_{R_l}, C_{R_{l_i}})> t_1} \wedge \xi(H_{R_l}, H_{R_{l_i}}) > t_2 $  where $\zeta$ and $\xi$ are two similarity functions and $t_1$ and $t_2$ are two user defined thresholds.
\end{definition} 

\subsubsection{Step-1: Identification of Similar Repair Locations}
\label{subsec:rlocations}
In order to find the evolutionary siblings ($E_{R_l}$) with respect to $R_l$, first we compute the similarity between $R_l$ and each location ($R_i$) in program spectra $S$. The intuition is that if the repair locations do not match, there is no reason for analyzing their context and version history. To this end, we apply Zhang and Shasha~\cite{zhang1989simple} tree distance algorithm to compute the similarity between two candidate repair locations $R_l$ and $R_i$. However, we introduce our own notion of similarity when comparing two AST nodes $N_1$ and $N_2$.

{\bf Kind Compatibility.} $N_1$ and $N_2$ are kind compatible if $kind(N_1)$ and $kind(N_2)$ have a super-class or sub-class relationship. For example, a {\tt variable access}, an {\tt array access}, or a {\tt method invocation} returning a value are of similar kind since all of them are expression type. However, a {\tt return} statement and a {\tt throw} statement are not kind compatible.

\textbf{Type Compatibility.} $N_1$ and $N_2$ are type compatible if $type(N_1)$ and $type(N_2)$ have either a super-class or sub-class relationship or satisfies the implicit type casting criteria defined by the language.

{\bf Name Similarity.} We extract the name of $N_1$ and $N_2$, and use {\it Levenshtein Distance} algorithm to compute textual similarity.

\sloppy It is worth noting that due to kind compatibility, similar but various kind of program constructs can be mapped to each other. For example, one developer may use {\tt area=length*width} whereas another may use {\tt area=table.getLength()*table.getWidth()}. Although a human can easily understand that {\tt length} and {\tt table.getLength()} represent similar data, in terms of ASTs, they are quite different. \tool{} is able to find such similarity, which is important in dealing with object oriented programming language. 

\begin{algorithm}
\caption{AST Node Similarity}

\label{alg:node_sim}
 \hspace*{\algorithmicindent} \textbf{Input:} Two AST Nodes $(N_1$ and $N_2)$\\
\hspace*{\algorithmicindent}\textbf{Output:} Similar($true/false$)?, Mapping of Elements $(M)$  
\begin{algorithmic}[1]
	\Statex \func{\underline{AstNodeSimilarity} }{$N_1$}{$N_2$}
		\State $s_1$ \IS{} \func{KindCompatibility}{$r_1$}{$r_2$}	
		\If{$s_1 == true$}
			\State $s_2$ \IS{} \func{TypeCompatibility}{$r_1$}{$r_2$}
			\If{$s_2 == true$}
				\State $s_3$ \IS{} \func{ComputeNameSim}{$r_1$}{$r_2$}
				\If{$s_3 > threshold$}
					\State \textbf{return} $true$;
				\EndIf
			\EndIf
		\EndIf
		\State \textbf{return} $false$;	
\end{algorithmic}
\end{algorithm}

Finally, we use Algorithm~\ref{alg:node_sim} to determine whether $N_1$ and $N_2$ are similar, and eventually Zhang-Shasha algorithm to determine whether $R_l$ and $R_i$ are similar. In the case of similar $R_l$ and $R_i$, we create a one-to-one node level mapping between the similar nodes in $R_l$ and $R_i$.

\subsubsection{Step-2: Determining the Relevance of Repair Locations}
\label{subsec:context}
Step-1 certainly would remove most of the irrelevant repair locations. However, there may be still a lot of false positives since any two random statements can be similar at a statement level, especially small statements. Therefore, \tool{} enhance the analysis with program context to make sure that the repair locations are indeed similar and used in a similar context. 

{\bf Significance of Program Context.} Program context (i.e., surrounding code around a target location) has been used in many software engineering tasks including program repair~\cite{Elixir:ASE2017}. To determine program context, the number of statements around the repair location is one of the important parameters. While a small context ($+1/-1$ statement) may not be sufficient to capture the developers' intent, a large context ($+/-10$ statements) may be difficult to generalize. To find evolutionary siblings, using adjacent program context is even further challenging since many of them are not clones in a traditional sense (as discussed in Section~\ref{sec:motivating_example}). 

{\bf Extraction of Program Context.} In order to overcome this challenge, unlike other repair tools~\cite{Elixir:ASE2017} that use a fixed size adjacent context, \tool{} uses a variable size non-contiguous relevant context (Definition~\ref{def:context}). More specifically, \tool{} uses {\it reaching-definition} based analysis within a method boundary to extract only relevant context, even though those statements are far away from the repair location. 
In compiler theory, a reaching definition for a given statement ($R_l$) is the closest earlier statement $R_i$ whose target variable can reach $R_l$ without an intervening assignment.

\begin{algorithm}
\caption{Extract Relevant Context}

\label{alg:context}
 \hspace*{\algorithmicindent} \textbf{Input:} A repair location ($R_l$), Source Code ($SC$)\\
\hspace*{\algorithmicindent}\textbf{Output:} Set of statements representing context ($C_{R_l}$) 
\begin{algorithmic}[1]
	\Statex \func{\underline{ExtractRelevantContext} }{$R_l$}{$SC$}
		\State $V$ \IS{}  \func{ExtractVariableAccesses}{$R_l$}
		\State $C_{R_l}$ \IS{} $\{R_l\}$
		\ForEach{$v$ in $V$}	
		\State $R_{l_v}$ \IS{} \func{ReachingDefinition}{$v$}
		\State $C_{R_l}$ \IS{} $C_{R_l} \cup \{R_{l_v}\}$
		\EndFor
		\If{$|C_{R_l}|$ = 1}
			\State $C_{R_l}$ \IS{} $C_{l_r} \cup$ \func{PreviousStatement}{$R_l$}
		\EndIf
		\State $C_{R_l}$ \IS{} \func{SortByLineNumber}{$C_{R_l}$}
		\State \textbf{return} $C_{R_l}$;
\end{algorithmic}
\end{algorithm}

We use Algorithm~\ref{alg:context} to extract the relevant non-contiguous repair context for a given repair location ($R_l$). In words, we first extract all the variable accesses from $R_l$. Then for each variable access, we determine the statement that satisfies the reaching definition property based on data-flow analysis. We take a union of all the statements obtained from the analysis sorted by the line number to form the context $C_{R_l}$. If $C_{R_l}$ contains only $R_l$, we add previous statement of $R_l$ in $C_{R_l}$.

{\bf Analysis of Program Context.} Once the program context for each repair location pair ($R_l, R_{l_i}$) is extracted, \tool{} applies the same tree matching algorithm from the previous step to determine whether the context are similar and the node mappings are still consistent. If they are similar, \tool{} marks them as potential evolutionary siblings. 

\subsubsection{Step-3: Revising Evolutionary Siblings Leveraging Version History}
\label{subsec:history}
Since the accuracy of identifying evolutionary siblings is a direct impact on the repair, we further leverage version history to revise them. There may be two potential scenarios. One, some repair locations in the candidate evolutionary siblings, identified in Step-2, are independent of each other. Two, some true evolutionary siblings are not in the list due to weak test specification. Our insight is that true evolutionary siblings may have a similar evolution history, i.e., they went through similar AST operations in the past. Therefore, version history may be helpful two mitigate both problems. 

Lets assume that we have three candidate evolutionary siblings ($R_{l_1}, R_{l_2}, R_{l_3}$). In order to revise the list with confidence, \tool{} first identifies all the commits ($C_1, C_2,..,C_n$) where the candidate repair locations were edited. Then it extracts the differences in each commit (before and after the changes) at the AST level (insertion, deletion, and modification). \tool{} further investigates each commit to identify if there are any other similar repair locations that have been also changed with the candidate siblings (applying Step-1). If \tool{} finds any similar repair locations ($R_4, R_5$), it analyzes their context as well (applying Step-2). If the results of both steps are positive, \tool{} adds such repair locations in the list of candidate evolutionary siblings. Therefore, for this example, \tool{} would get five candidate siblings. However, even if \tool{} adds some plausible siblings in this phase, it can safely  discard them during validation (Section~\ref{validation}), if they introduce any regression failure, to eventually generate a correct patch.  Finally, \tool{} removes the repair locations that do not have the similar evolution history based on the edit operations. \tool{} passes the final evolutionary siblings along with the mapping of similar nodes among the repair locations to the next step for the generation of candidate patches. 

It should be noted that in the whole process of finding evolutionary siblings, \tool{} applies AST matching algorithm lazily in several steps. It is indeed possible to find all the evolutionary siblings in the entire code-base and then analyze their version history. However, this approach would be inefficient. 

\subsection{Generation of Candidate Patches}

In this step, \tool{} generates a single or multi-hunk patch depending on the results from the previous step. For a single repair location ($R_l$), \tool{} follows the traditional patch generation approach, i.e., it instantiates a select repair schema ($\Im$) with the repair expressions in scope. However, for multiple repair locations ($\Re=\{R_l, R_{l_1},..,R_{l_n}\}$), where evolutionary siblings are involved, \tool{} instantiates the selected repair schema at an abstract level. More specifically, given a group of repair locations ($\Re$) and a mapping between similar AST nodes in $\Re$, denoted by $M(\Re)$, \tool{} abstracts all repair locations to remove the differences due to various identifier names. Then the same repair schema, $\Im$ is instantiated at the abstract level at each repair location simultaneously. Once the transformation, $\Im$  is applied, \tool{} generates the concrete candidate patches by reverting the original variables using $M(\Re)$. Therefore, the repair space generated by \tool{} may have both single or multi-hunk patches. It is worth noting that, due to simultaneous repair schema instantiation, \tool{} keeps the repair space comparable to the repair tools that only focuses on generating single-hunk patch.

\subsection{Ranking of Candidate Patches and Validation}
\label{validation}

Once the candidate patches are generated, \tool{} can use any ranking models (heuristic based or machine learning based) proposed in the existing repair tools to rank the candidate patches, and selects Top $N$ candidate patches for validation, one at a time. During the validation phase for each candidate (single or multi-hunk) patch, \tool{} first runs the failing tests, and if they pass, \tool{} runs the regression tests. If the regression test suite pass, \tool{} stops and reports that patch as a final patch.
\section{Experimental Setup}
\label{sec:exp_setup}

\subsection{Implementation} 
{\bf \tool} is a G\&V-style repair tool implemented in Java. It includes a spectrum-based fault localizer as well as a source code and version history analyzer. \tool{} also implements a mechanism for applying repair schemas simultaneously at several locations, to effect multi-hunk repairs. 
Inspired by previous well-known program repair tools~\cite{PAR:ICSE2013, ACS:ICSE2017, Elixir:ASE2017}, \tool{} includes repair schemas such as checking null pointer check, changing and inserting method invocation, changing and inserting {\it if} conditions, and so on.
Further, it incorporates a machine learning based patch ranking model similar to \elixir~\cite{Elixir:ASE2017}. 
For the details of each of the repair schemas and the ranking model of candidate patches, the reader is referred to the corresponding papers~\cite{PAR:ICSE2013, ACS:ICSE2017, Elixir:ASE2017}.
We also implement several baseline versions of \tool{} to demonstrate the effectiveness of its various components (Section~\ref{contribution-of-components}).

\subsection{Dataset}

We used the popular \dfj dataset~\cite{just2014defects4j} to evaluate \tool{}, specifically the five subjects: Math, Lang, Chart, Time, and Closure.

\subsection{Training \tool{}}

Since \tool{} uses a machine learning technique to rank and prune candidate patches for validation, we train \tool{} with real-world bugs. In order to train \tool{}, we used another publicly available real-world bug dataset, \bugsjar. There are 1,158 real-world bugs in \bugsjar taken from eight well-known large Apache projects. Among them, Apache Commons Math is common to both \dfj and \bugsjar~\cite{saha2018bugs}. Therefore, we removed Apache Commons Math from our training set, to keep the training and testing datasets mutually exclusive. 

\subsection{Research Questions}

\begin{itemize}[noitemsep,topsep=1pt,parsep=1pt,partopsep=1pt,leftmargin=24pt]

\item[\textbf{RQ1:}] How effective is \tool{} for repair localization through evolutionary sibling detection compared to traditional clone detection?

\item[\textbf{RQ2:}] How effectively does \tool{} generalize the traditional single-hunk repair strategy of APR tools to perform both single-hunk as well as multi-hunk repair?

\item[\textbf{RQ3:}] How effective is \tool{} in repairing programs compared to state-of-the-art program repair tools?

\item[\textbf{RQ4:}] What is the contribution of various components in \tool{} to its overall bug-fixing capability?
\end{itemize}

\subsection{Experimental Configurations}

We ran all the experiments on a cluster of Virtual Machines (VM), where each VM was configured to have double core 3.6GHz processor and 4GB memory. We used Ubuntu 16.04 LTS operating system and Java 7. There are several configuration parameters in \tool{} as well. \tool{} used threshold value 0.8 for determining tree similarity, iterated through Top 200 repair locations and selected 50 candidate patches per repair schema.  We set a time out of 5 hours following the recently introduced repair tool, SimFix~\cite{SimFix:ISSTA2018}.

\section{Results}
\label{sec:results}

\subsection{RQ1: Effectiveness of  Repair Localization}
\label{repair-localization}

{\bf Motivation.} The main contribution of \tool{} is that it enables fixing a specific but prominent class of multi-hunk bugs by accurate repair localization, which is in turn achieved through the detection of evolutionary siblings. Therefore, it is important to evaluate how effective \tool{} is in detecting evolutionary siblings, especially compared to a traditional clone detection tool since if a clone detector can detect all the evolutionary siblings accurately we do not need any sophisticated repair localization. 

{\bf Experiment.} In order to make a meaningful comparison, we first identify all the relevant bugs in \dfj{} for this experiment. Here, by {\it relevant bugs} we mean all the bugs that involve either only a single-hunk patch or ones with a multi-hunk patch but similar edits in all hunks. The rest of the bugs are, by definition, out of scope and will only add noise to the experimental results. 
This gave us 154 bugs in total, 130 single-hunk bugs and 24 multi-hunk bugs. Recall from Section~\ref{subsec:siblings} that, for a given buggy location, the objective of repair localization is to identify all the relevant locations where similar changes are required to repair the bug correctly and completely. Therefore, in an ideal case, if we point a clone detector or \tool{} to an actual buggy line (\ie an input location), it should return only one repair location (the input location itself) for a single-hunk bug and all the $n$ relevant repair locations (including the input location) for an $n$-hunk bug. To this end,  we ran the {\it repair localization} component of \tool{} and an established clone detector, Deckard~\cite{deckard:ICSE2007} on all the 154 bugs in \dfj with respect to their actual buggy locations (or the one with the highest fault-localization rank, for a multi-hunk bug). For Deckard, we set the minimum number of tokens to 10 (approximately two lines of code) following~\cite{LASE:ICSE2013}. 

\begin{table}
	\setlength{\tabcolsep}{4pt}
	\begin{center}
	\captionsetup{justification=centering}
		\caption{Effectiveness of \tool and Deckard for \\Repair Localization}
		\label{tbl:repair-localization} 
		\begin{tabular}{lrlrr}
			\toprule
			{\bf Patch-Type}&{\bf Bugs}&{\bf Approach}&{\bf Correct} & {\bf Proportion} \\\midrule
			Single-hunk&130& \tool{}	& 114 & 88\%	\\
			&&Deckard& 38 &	29\%	\\\hline
			Multi-hunk & 24&\tool{}&15 & 63\%\\
			&&Deckard& 2 &	8\%	\\\hline
			Total& 154&\tool{}&129 & 84\%\\
			&&Deckard& 40 &	26\%	\\
			\bottomrule
		\end{tabular}
	\end{center} 
	\vspace{0pt}
\end{table}

{\bf Results.} From Table~\ref{tbl:repair-localization}, we see that \tool{} identified the repair locations correctly for 84\% (129 out of 154) bugs whereas Deckard found the correct locations only for 26\% (40 out of 154) bugs. On closer inspection, we see that Deckard performed even poorer on multi-hunk bugs, which is the main target of repair localization, than single-hunk bugs. It detected correct repair locations for only 2 bugs, while \tool{} correctly localized 15 out of 24 (63\%) of these bugs. Even for single-hunk bugs, Deckard correctly localized only 29\% (38/130) of the instances compared to \tool{}'s 88\%. In both single-hunk and multi-hunk instances Deckard produced false positives as well as false-negatives.

The reason is simply that many of the target repair locations are not clones in a traditional sense (like our motivating example in Figure~\ref{fig:math-24}), and beyond the scope of Deckard's purely syntactic, fixed-context clone detection. By contrast, \tool{}'s specialized evolutionary sibling analysis, which combines several orthgonal sources of information, is far more accurate, and hence indispensible, for the application at hand, \ie multi-hunk program repair.

\subsection{RQ2: Generalizing Single-hunk Repair}

{\bf Motivation.} In RQ1, we demonstrated that \tool{} is effective in detecting evolutionary siblings. The current experiment evaluates how well this repair localization translates into the end goal of producing correct patches, not just for multi-hunk instances but also single-hunk patches.

{\bf Experiment.} We first create a baseline, named \toolsh{} that has all the repair schemas in \tool{} but only performs single hunk repair. Then we ran both \tool{} and \toolsh{} on all the 154 bugs described in RQ1.

\begin{table}
	\vspace{-1pt}
	\setlength{\tabcolsep}{3pt}
	\begin{center}
		\caption{Effectiveness of \tool (Correct/Incorrect)}
		\vspace{-5pt}
		\label{tbl:effectiveness} 
		\begin{tabular}{lcccccc}
			\toprule
			{\bf Subject}&{\bf Math}&\bf {Lang}&{\bf Time}&{\bf Chart}&{\bf Closure}&{\bf Total}\\\midrule
			\tool	  & 20/9 & 10/5	& 3/2 &	8/2	& 8/5 &	\numfixedbugs/\numincorrectfixedbugs\\
			\toolsh{} & 12/9 &  8/3 & 2/2 & 6/2 & 6/5 & \numsinglebugs/\numincorrectsinglebugs\\
			\bottomrule
		\end{tabular}
	\end{center} 
	\vspace{-1pt}
\end{table}

{\bf Results.} As presented in Table~\ref{tbl:effectiveness}, \tool{} generates \numfixedbugs correct patches and \numincorrectfixedbugs incorrect patches while \toolsh{} generates \numsinglebugs correct patches and \numincorrectsinglebugs incorrect patches as well. A deeper look into the results reveals that \tool{} fixed 34 single-hunk bugs and \nummultibugs multi-hunk bugs. By comparing the results with \toolsh{}, we observe that \tool{} lost no single-hunk bug due to incorrect simultaneous repair. Moreover, \tool{} generates only two additional incorrect patches due to simultaneous multi-hunk repair. The reason is that in multi-hunk repair, since program changes happen in two or more repair locations, the probability of detecting regression bugs by the test suite increases significantly. Even if a test case detects regression in one repair location out of $n$, the whole candidate patch is discarded.

\begin{figure}
\begin{center}
\includegraphics[width=0.95\columnwidth]{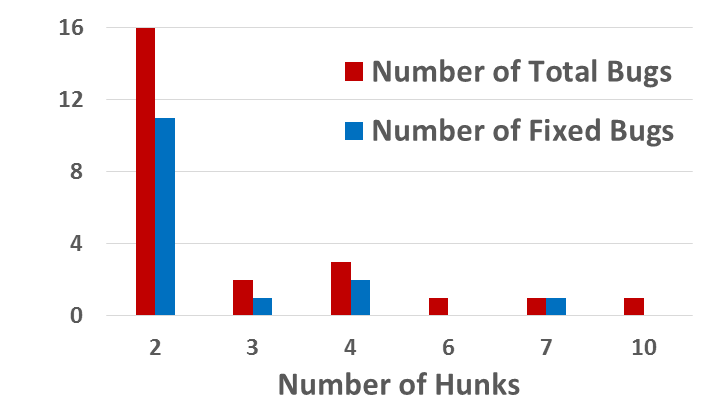}
\end{center}
\caption{Distribution of Multi-Hunk Bugs in Our Study}
\label{fig:hunk}
\end{figure}

An analysis of the distribution of multi-hunk bugs and patches, in terms of number of hunks, offers some interesting observations. As presented in Figure~\ref{fig:hunk}, \tool{} fixed a variety of multi-hunk bugs, ranging in size from 2 hunks to as large as 7 hunks.

\subsection{RQ3: \tool{} vs. State of the art}
\label{state-of-the-art-comparison}

{\bf Motivation.} In RQ2, we demonstrated that \tool{} is effective in fixing multi-hunk bugs as well as single-hunk bugs. In this section, we evaluate the effectiveness of \tool{} with respect to the state-of-the-art program repair approaches. 

{\bf Experiment.} We choose six of the most recent (Java) APR tools: SimFix~\cite{SimFix:ISSTA2018}, CapGen~\cite{CapGen:ICSE2018}, JAID~\cite{JAID:ASE2017}, \elixir~\cite{Elixir:ASE2017}, ssFix~\cite{ssFix:ASE2017}, and ACS~\citep{ACS:ICSE2017} as representative of the state of the art. Since all these tools have already been evaluated on \dfj, we simply take the results from the respective papers. 

{\bf Results.} Table~\ref{tbl:comparison} presents the number of correct and incorrect patches generated by various tools. From the results, we see that \tool generated \numfixedbugs correct patches, which is the most among all tools. It fixes 15 more patches than the next best tool, SimFix. In Table~\ref{tbl:results_by_bug_id_sh} and Table~\ref{tbl:results_by_bug_id_mh}, we provide more detailed results at the individual bug level. More specifically, Table~\ref{tbl:results_by_bug_id_sh} and Table~\ref{tbl:results_by_bug_id_mh} present all the single and multi-hunk bug IDs in Defects4J respectively, for which at least one tool generated a correct patch. For each table, a check-mark (\checkmark) and a cross (\ding{55}) represent a correct patch and an incorrect patch respectively. A hyphen (-) denotes that the corresponding tool did not generate any patch for that bug ID. Finally, the ^^21 sign denotes that we are not aware of the results reported by the tool.

Since SimFix and ACS have also the ability to fix multi-hunk bugs, we investigated the results at the level of individual bugs. We found that \tool{} fixes \numnewmultibugs unique multi-hunk bugs that SimFix and ACS cannot fix. Furthermore, if we consider all the bugs, \tool{} fixed \numnewbugs unique bugs that no existing approaches could fix. Overall the results demonstrate that \tool{} advances the state of the art significantly.

\begin{table}
\vspace{-1pt}
\setlength{\tabcolsep}{4pt}
\begin{center}
\caption{Statistics of Patch Generation by Various Techniques (Correct/Incorrect)}
\vspace{-5pt}
\label{tbl:comparison} 
\begin{tabular}{lcccccc}
\toprule
{\bf Subject}&{\bf Math}&\bf {Lang}&{\bf Time}&{\bf Chart}&{\bf Closure}&{\bf Total}\\\midrule
\tool		& 20/9	&	10/5	&	3/2	&	8/2	& 8/5 &	\cellcolor{gray!25} \numfixedbugs/\numincorrectfixedbugs\\
SimFix		& 14/12	&	9/4		&	1/0	& 	4/4	& 6/2 & 34/22 \\
CapGen		& 13/^^21 & 5/^^21 & 0/^^21 & 4/^^21 & -/- & 22/^^21\\
JAID		& 1/^^21 & 1/^^21 & 0/^^21 & 2/^^21 & 5/^^21 & 9/^^21\\
\elixir		& 12/7	&	8/4	&	2/1	&	4/3	& -/- &	26/15\\
ssFix		& 10/16	& 5/7	& 0/4	& 3/4	& 2/9 & 20/40\\
ACS			&	12/4		&	3/1	&	1/0	& 2/0 & -/-	&	18/5\\
\bottomrule
\end{tabular}
\end{center} 
\vspace{0pt}
\end{table}

\begin{table}
\vspace{-1pt}
\setlength{\tabcolsep}{2pt}
\begin{center}
\caption{Results for Single-hunk Bug IDs}
\vspace{-5pt}
\label{tbl:results_by_bug_id_sh} 
\begin{tabular}{lcccccc}
\toprule
{\bf Bug ID}&{\bf \tool}&\bf {SimFix}&{\bf CapGen}&{\bf ssFix}&{\bf \elixir}&{\bf ACS}\\\midrule
Chart-1	& \checkmark & \checkmark & \checkmark & \checkmark & \checkmark & -\\ 
Chart-3	& \checkmark & \checkmark & ^^21 & - & \ding{55} & -\\
Chart-8	& \checkmark & - & \checkmark & - & \checkmark & -\\ 
Chart-9	& \checkmark & - & ^^21 & \ding{55} & \checkmark & -\\ 
Chart-11 & \checkmark & - & \checkmark & - & \checkmark & -\\ 
Chart-12 & \checkmark & \ding{55} & ^^21 & - & - & -\\  
Chart-20 & - & \checkmark & ^^21 & \checkmark & - & -\\
Chart-24 & - & - &  \checkmark & \checkmark & - & -\\
Closure-14 & \checkmark & \checkmark & ^^21 & \checkmark & - & -\\  
Closure-57	& - & \checkmark & ^^21 & - & - & -\\ 
Closure-62* & \checkmark & \checkmark & ^^21 & - & - & -\\ 
Closure-63* & \checkmark & \checkmark & ^^21 & - & - & -\\ 
Closure-73 & \checkmark & \checkmark & ^^21 & - & - & -\\ 
Closure-86 & \checkmark & - & ^^21 & - & - & -\\ 
Closure-92 & \checkmark & - & ^^21 & - & - & -\\
Closure-109	& \checkmark & - & ^^21 & \ding{55} & - & -\\ 
Lang-6	& \checkmark & - & \checkmark & \checkmark & \checkmark & -\\ 
Lang-16 & - & \checkmark & \checkmark & - & - & -\\
Lang-21	& - & - & ^^21 & \checkmark & - & -\\ 
Lang-24	& \checkmark & - & ^^21 & - & \checkmark & \checkmark\\ 
Lang-26	& \checkmark & - & ^^21 & - & \checkmark & -\\
Lang-33	& \checkmark & \checkmark & ^^21 & \checkmark & \checkmark & -\\ 
Lang-38	& \checkmark & - & ^^21 & - & \checkmark & -\\  
Lang-39 & \ding{55} & \checkmark & ^^21 & \ding{55} & \ding{55} & \ding{55}\\ 
Lang-43	& \checkmark & \checkmark & \checkmark & \checkmark & \checkmark & -\\ 
Lang-57	& \checkmark & - & \checkmark & - & \checkmark & -\\  
Lang-58 & \ding{55} & \checkmark & ^^21 & \ding{55} & \ding{55} & -\\
Lang-59	& \checkmark & - & \checkmark & \checkmark & \checkmark & -\\   
Math-3	& - & - & ^^21 & - & - & \checkmark\\
Math-5	& \checkmark & \checkmark & \checkmark & - & \checkmark & \checkmark\\
Math-25	& \checkmark & - & ^^21 & - & - & \checkmark\\ 
Math-30	& \checkmark & - & \checkmark & \checkmark & \checkmark & -\\ 
Math-33	& \checkmark & \checkmark & \checkmark & \checkmark & \checkmark & -\\ 
Math-34	& \checkmark & - & ^^21 & - & \checkmark & -\\ 
Math-41 & - & \checkmark & ^^21 & \checkmark & - & -\\ 
Math-50 & \ding{55} & \checkmark & ^^21 & \checkmark & \checkmark & -\\ 
Math-53 & - & \checkmark & \checkmark & \checkmark & - & -\\
Math-57	& \checkmark & \checkmark & \checkmark & \checkmark & \checkmark & -\\ 
Math-58	& \checkmark & - & \checkmark & - & \checkmark & -\\ 
Math-59	& \checkmark & \checkmark & \checkmark & \checkmark & \checkmark & -\\  
Math-61	& - & - & ^^21 & - & - & \checkmark\\
Math-63 & \ding{55} & \checkmark & \checkmark & - & \ding{55} & -\\
Math-70	& \checkmark & \checkmark & \checkmark & \checkmark & \checkmark & -\\
Math-75	& \checkmark & \checkmark &  \checkmark & - & \checkmark & -\\ 
Math-80 & \ding{55} & \ding{55} & \checkmark & \checkmark & \ding{55} & -\\
Math-82	& \checkmark & \ding{55} & ^^21 & - & \checkmark & \checkmark\\ 
Math-85	& \checkmark & \ding{55} & \checkmark & \ding{55} & \checkmark & \checkmark\\ 
Math-89	& - & - & ^^21 & - & - & \checkmark\\ 
Time-4	& \checkmark & - & ^^21 & \ding{55} & \checkmark & -\\
Time-7 & - & \checkmark & ^^21 & - & - & -\\
Time-15	& \checkmark & - & ^^21 & - & \checkmark & \checkmark\\ 
\bottomrule
\end{tabular}
* Closure-62 and Closure-63 are duplicate bugs. We removed any duplication from \tool's bug count.
\end{center} 
\vspace{0pt}
\end{table}

\begin{table}
\vspace{-1pt}
\setlength{\tabcolsep}{2pt}
\begin{center}
\caption{Results for Multi-hunk Bug IDs}
\vspace{-5pt}
\label{tbl:results_by_bug_id_mh} 
\begin{tabular}{lcccccc}
\toprule
{\bf Bug ID}&{\bf \tool}&\bf {SimFix}&{\bf CapGen}&{\bf ssFix}&{\bf \elixir}&{\bf ACS}\\\midrule
Chart-7	& - & \checkmark & ^^21 & \ding{55} & - & -\\ 
Chart-14 & \checkmark & \ding{55} & ^^21 & - & - & \checkmark\\ 
Chart-19 & \checkmark & - & ^^21 & - & - & \checkmark\\ 
Closure-4 & \checkmark & - & ^^21 & - & - & -\\
Closure-78 & \checkmark & - & ^^21 & - & - & -\\ 
Closure-115 & - & \checkmark & ^^21 & \checkmark & - & -\\
Lang-7	& - & - & ^^21 & - & - & \checkmark\\
Lang-27 & - & \checkmark & ^^21 & \ding{55} & - & -\\
Lang-35	& - & - & ^^21 & - & - & \checkmark\\ 
Lang-41 & - & \checkmark & ^^21 & - & - & -\\
Lang-47	& \checkmark & - & ^^21 & - & - & -\\  
Lang-50 & - & \checkmark & ^^21 & - & - & -\\
Lang-60	& \checkmark & \checkmark & ^^21 & - & - & -\\ 
Math-4	& \checkmark & - & ^^21 & - & - & \checkmark\\ 
Math-24	& \checkmark & - & ^^21 & - & - & -\\ 
Math-35	& \checkmark & \checkmark & ^^21 & - & - & \checkmark\\
Math-43	& \checkmark & - & ^^21 & - & - & -\\ 
Math-46	& \checkmark & - & ^^21 & - & - & -\\ 
Math-49	& \checkmark & - & ^^21 & - & - & -\\  
Math-65 & - & - &  \checkmark & - & - & -\\
Math-71 & - & \checkmark & ^^21 & - & - & -\\ 
Math-72	& \checkmark & \ding{55} & ^^21 & - & - & -\\ 
Math-79 & - & \checkmark & ^^21 & \checkmark & - & -\\
Math-90	& - & - & ^^21 & - & - & \checkmark\\ 
Math-93	& - & - & ^^21 & - & - & \checkmark\\ 
Math-98	& \checkmark & \checkmark & ^^21 & - & - & -\\
Math-99	& - & - & ^^21 & - & - & \checkmark\\
Time-26	& \checkmark & - & ^^21 & - & - & -\\ 
\bottomrule
\end{tabular}
\end{center} 
\vspace{0pt}
\end{table}

\subsection{RQ4: Contribution of Various Components of \tool}
\label{contribution-of-components}

{\bf Motivation.} Evaluations in the previous RQs demonstrate that \tool{} overall outperforms existing tools. The next experiment evaluates the contribution of various features of \tool{} to its overall bug-fixing capability.

{\bf Experiment.} In order to investigate the contribution of various components, we create three versions of \tool{}.

\begin{itemize}
\item \toolc uses a fixed number of lines $w$ before and after the buggy location as its context, rather than the semantic context. For this version we experimented with  $ w = \{1, 2, 3\}$ and reported the best results.

\item \toolh ignores the version history in repair.

\item \tooli implements an incremental repair strategy where there is no notion of evolutionary siblings. Each location is fixed independently, one after another, as long as each successive repair decreases the number of failing tests. This strategy broadly mimics the one implemented in SimFix~\cite{SimFix:ISSTA2018} and ACS~\cite{ACS:ICSE2017}.
\end{itemize} 

We run this experiment only for the \numfixedbugs bugs that \tool{} correctly fixed. 

{\bf Results.} Table~\ref{tbl:comparison_among_variants} presents the aggregated results in terms of number of correct patches. From the results, we observe that when we used a fixed size context, \tool{} lost 11 bugs. Similarly, version history was crucial in fixing six bugs correctly. Finally, the results further show that when \tool{} does not leverage the information of evolutionary siblings, \ie works in an incremental fashion, it cannot fix 9 bugs. In summary, the results demonstrate that all the features of \tool{} contributed in fixing multi-hunk bugs.

\begin{table}
	\setlength{\tabcolsep}{2pt}
	\begin{center}
		\caption{Comparison among variants of \tool{}}
		\vspace{-5pt}
		\label{tbl:comparison_among_variants} 
		\begin{tabular}{lc}
			\toprule
			{\bf Approach}&{\bf Number of Correct Patches}\\\midrule
			\tool{}		& \numfixedbugs\\
			\toolc	& 38\\
			\toolh	& 43\\
			\tooli{}	& 40\\
			\bottomrule
		\end{tabular}
	\end{center} 
	\vspace{0pt}
\end{table}

\section{Limitations \& Threats to Validity}
\label{sec:discussion}

\textbf{\textit{Scope of multi-hunk repairs.}} Our current technique addresses only a specific class of multi-hunk repairs, namely ones with substantially similar patches for each hunk. While this does boost the successful repairs by almost 50\% compared to the baseline tool, future research needs to address other classes of multi-hunk bugs, the vast majority of which are still out of scope for \tool{} or any other APR tool.

\textbf{\textit{Accuracy of version history analysis.}} Our version history analysis can be impacted by noise in the revision history introduced by major structural changes to the repository, such as to the directory or package structure, or other systemic re-factoring changes. We use several heuristics to compensate for such disruptions and manually inspected the history of a few sampled bug instances to verify the accuracy of the heuristics. However, we cannot guarantee the soundness of the analysis.

\textbf{\textit{Generalizability of the results.}} Our evaluation was only carried out on the Defects4J dataset, which is a widely used benchmark for program repair research. However, the dataset's 5 subject systems cannot capture the wide variety of Java applications and their bugs. Further validation of this technique on other subjects should necessarily be done in future. Further, our current repair results depend on the capabilities of the baseline repair tool, on which our multi-hunk repair technique is implemented. Although our baseline APR tool is quite competitive with the state of the art, using a different or improved APR tool, such as SimFix~\cite{SimFix:ISSTA2018} for example, could change or improve the results. Lastly, our technique has been instantiated for Java program repair but could, in principle, be applied to C/C++ program repair as well. But its efficacy in that setting remains to be investigated.

\section{Related Work}
\label{sec:related_work}

\textbf{Automatic program repair.} A decade of research has generated a rich body of work on program repair, summarized in two excellent recent surveys~\cite{APRSurvey:2018, Monperrus:2018}. With the notable exception of Angelix~\cite{Angelix:ICSE2016} APR research so far does not target multi-location bugs, which is our main focus. However, APR research most related to \tool's approach can be classified into techniques that: (1) mine existing code or patches for repair fragments, \eg variables, expressions, statements, or complete code snippets, \etc, (2) learn abstract repair spaces, \eg program transformations from existing patches, and (3) can produce multi-hunk patches.

\textbf{APR - Mining repair fragments:} GenProg~\cite{GenProg:ICSE2012}, which pioneered this area, uses genetic search on a space of repair mutations formed by code snippets copied from elsewhere in the program. RSRepair~\cite{RSRepair:ICSE2014} and AE~\cite{AE:ASE2013} follow GenProg, using random and deterministic search respectively, instead. $\mu$\textsc{Scalpel}~\cite{muScalpel:ISSTA2015} transplants code snippets mined from a donor application to repair bugs in a donee application. CodePhage~\cite{CodePhage:PLDI2015} also performs transplantation but targets missing if-condition related bugs. ACS~\cite{ACS:ICSE2017} performs if condition repairs using predicates mined from Github. SearchRepair~\cite{SearchRepair:ASE2015} mines repair fragments using a semantic search, based on SMT formulas and constraint solving. ssFix~\cite{ssFix:ASE2017} performs the same search using a general purpose search engine. SimFix~\cite{SimFix:ISSTA2018} also searches for a donor snippet, but further uses a mined space of abstract schemas to prune the search.
Prophet~\cite{Prophet:POPL2016} and Elixir~\cite{Elixir:ASE2017} do not directly mine repair artifacts but rather use a corpus of existing patches to train a classifier, which is then used to rank the space of concrete patches. As a whole the above techniques search for compatible code fragments (or features thereof) to contribute to the repair of the bug at hand \ie a donor-donee relationship. By contrast, \tool{}'s search for ``similar" code is used to find a set of evolutionary siblings that can be repaired concurrently. 

\textbf{APR - Learning abstract repair spaces:} PAR~\cite{PAR:ICSE2013} first used this approach, defining its repair space using a set of $10$ specialized repair templates manually derived from human-written patches. Relifix~\cite{Relifix:ICSE2015} uses specialized repair schemas customized for software regression errors. History-driven repair~\cite{HDRepair:SANER2016} prioritizes its pool of candidate repairs based on the frequency of occurrence of the repair in a corpus of past (human-written) patches. Tan et al.~\cite{antiPatterns:FSE2016} propose a blacklist abstract repair space defined by a set of anti-patterns, to curb the generation of incorrect patches.
Genesis~\cite{Genesis:FSE2017} automatically extracts a set of repair schemas for specific classes of bugs by solving an optimization problem on the set of previous patches for each bug class.
CapGen~\cite{CapGen:ICSE2018} mines a set of 30 frequently occurring AST-level transformations from a corpus of previous patches, and uses them for repair. Our contribution is fundamentally orthogonal to the above body of work in that our identification of evolutionary siblings seeks to identify \emph{where} the repair should be performed while the above informs \emph{how} the repair at a \emph{given location} should be performed.

\textbf{APR - Multi-hunk repair:} So far Angelix~\cite{Angelix:ICSE2016} is the only APR tool to specifically target multi-hunk patches. It generates a symbolic oracle for potential changes to the top `k' locations given by fault localization and then independently synthesizes patches for each location from the oracle. The oracle may implicitly capture any inter-location dependencies reflected in the test suite. By contrast, \tool{} uses multiple sources of information to pro-actively derive and exploit sibling relationships between repair locations.
Although not specifically described in \cite{ACS:ICSE2017, SimFix:ISSTA2018} the implementations of ACS~\cite{ACS:ICSE2017}  and SimFix~\cite{SimFix:ISSTA2018} are in fact capable of performing simple instances of multi-hunk repair. Specifically, when there are distinct test-cases separately implicating each of many bug locations, the tool independently repairs these bug locations, one after another, using the \emph{number} of passing tests as a progress metric. Our key contribution is that we employ sibling relationships between locations to significantly cut down the search space by repairing locations simultaneously, localize the repair more accurately, and can even repair locations not directly implicated by the test spectrum.

\textbf{Code Clone Detection.} Our work is broadly inspired by research on code clone detection and extraction of code clone genealogies. Deckard~\cite{deckard:ICSE2007}, NiCad~\cite{NiCad:ICPC2011}, and CCFinder~\cite{CCFinder:TSE2002} are some of the popular clone detectors.  Kim et al.~\cite{kim2005using} were the first to extracted and analyze clone genealogies across multiple revisions of a software system to understand the evolution of code clones. However, as demonstrated in RQ3, the evolutionary siblings that form the basis of our approach may or may not be traditional code clones.

\textbf{Systematic Edits.} Our approach is also inspired by research on systematic edits~\cite{Kim:ICSE2009, Meng:PLDI2011}, with LASE~\cite{LASE:ICSE2013} and \textsc{Rase}~\cite{Rase:ICSE2015} being the most advanced tools in this area. The key idea in this research is to learn a common abstract edit script from a few examples instances of it, which is then replicated at target locations in the code, also identified using the learned edit script. \tool{} also shares the notion of similar edits at multiple locations. However, we do not rely on examples for identifying sibling locations.

\section{Conclusion}
\label{sec:conclusion}

Automatic program repair techniques have made significant advances over the past decade. However, practical deployment remains an elusive goal. One of the significant obstacles to achieving this goal, is the inability of current APR techniques to produce multi-hunk patches. In this work, we presented a novel APR technique that generalizes single-hunk repair to encompass a specific but significant class of multi-hunk repair problems, namely ones that require applying a substantially similar patch at a number of locations. We term such sets of repair locations as evolutionary
siblings -- similar looking code, instantiated in similar contexts, that are expected to undergo similar changes over time. We proposed a novel analysis to accurately identify a set of evolutionary siblings, for a given bug. This analysis combines three orthogonal sources of information, namely, the test-suite spectrum, a novel code similarity analysis that compares both syntactic and semantic features, and the revision history of the project. We implemented this technique in a tool \tool{} and demonstrated that it is able to correctly fix \numfixedbugs bugs in the Defects4J dataset, the highest of any individual APR technique to date. This includes \nummultibugs multi-hunk bugs, and \numnewbugs bugs which have not been fixed by any other technique so far. We see this contribution as a small but important step on the road to achieving practical deployment of APR tools.

\bibliographystyle{ACM-Reference-Format}
\bibliography{references} 


\begin{thebibliography}{48}


\ifx \showCODEN    \undefined \def \showCODEN     #1{\unskip}     \fi
\ifx \showDOI      \undefined \def \showDOI       #1{#1}\fi
\ifx \showISBNx    \undefined \def \showISBNx     #1{\unskip}     \fi
\ifx \showISBNxiii \undefined \def \showISBNxiii  #1{\unskip}     \fi
\ifx \showISSN     \undefined \def \showISSN      #1{\unskip}     \fi
\ifx \showLCCN     \undefined \def \showLCCN      #1{\unskip}     \fi
\ifx \shownote     \undefined \def \shownote      #1{#1}          \fi
\ifx \showarticletitle \undefined \def \showarticletitle #1{#1}   \fi
\ifx \showURL      \undefined \def \showURL       {\relax}        \fi
\providecommand\bibfield[2]{#2}
\providecommand\bibinfo[2]{#2}
\providecommand\natexlab[1]{#1}
\providecommand\showeprint[2][]{arXiv:#2}

\bibitem[\protect\citeauthoryear{Abreu, Zoeteweij, Golsteijn, and van
  Gemund}{Abreu et~al\mbox{.}}{2009}]%
        {Abreu:2009}
\bibfield{author}{\bibinfo{person}{Rui Abreu}, \bibinfo{person}{Peter
  Zoeteweij}, \bibinfo{person}{Rob Golsteijn}, {and} \bibinfo{person}{Arjan
  J.~C. van Gemund}.} \bibinfo{year}{2009}\natexlab{}.
\newblock \showarticletitle{A Practical Evaluation of Spectrum-based Fault
  Localization}.
\newblock \bibinfo{journal}{\emph{Journal of Systems and Software}}
  \bibinfo{volume}{82}, \bibinfo{number}{11} (\bibinfo{date}{Nov.}
  \bibinfo{year}{2009}), \bibinfo{pages}{1780--1792}.
\newblock


\bibitem[\protect\citeauthoryear{Barr, Brun, Devanbu, Harman, and Sarro}{Barr
  et~al\mbox{.}}{2014}]%
        {barr2014plastic}
\bibfield{author}{\bibinfo{person}{Earl~T Barr}, \bibinfo{person}{Yuriy Brun},
  \bibinfo{person}{Premkumar Devanbu}, \bibinfo{person}{Mark Harman}, {and}
  \bibinfo{person}{Federica Sarro}.} \bibinfo{year}{2014}\natexlab{}.
\newblock \showarticletitle{The plastic surgery hypothesis}. In
  \bibinfo{booktitle}{\emph{Proceedings of the 22nd ACM SIGSOFT International
  Symposium on Foundations of Software Engineering}}. ACM,
  \bibinfo{pages}{306--317}.
\newblock


\bibitem[\protect\citeauthoryear{Barr, Harman, Jia, Marginean, and Petke}{Barr
  et~al\mbox{.}}{2015}]%
        {muScalpel:ISSTA2015}
\bibfield{author}{\bibinfo{person}{Earl~T. Barr}, \bibinfo{person}{Mark
  Harman}, \bibinfo{person}{Yue Jia}, \bibinfo{person}{Alexandru Marginean},
  {and} \bibinfo{person}{Justyna Petke}.} \bibinfo{year}{2015}\natexlab{}.
\newblock \showarticletitle{Automated Software Transplantation}. In
  \bibinfo{booktitle}{\emph{Proceedings of the 2015 International Symposium on
  Software Testing and Analysis}} \emph{(\bibinfo{series}{ISSTA 2015})}.
  \bibinfo{publisher}{ACM}, \bibinfo{address}{New York, NY, USA},
  \bibinfo{pages}{257--269}.
\newblock


\bibitem[\protect\citeauthoryear{Chen, Pei, and Furia}{Chen
  et~al\mbox{.}}{2017}]%
        {JAID:ASE2017}
\bibfield{author}{\bibinfo{person}{Liushan Chen}, \bibinfo{person}{Yu Pei},
  {and} \bibinfo{person}{Carlo~A Furia}.} \bibinfo{year}{2017}\natexlab{}.
\newblock \showarticletitle{Contract-based program repair without the
  contracts}. In \bibinfo{booktitle}{\emph{Automated Software Engineering
  (ASE), 2017 32nd IEEE/ACM International Conference on}}. IEEE,
  \bibinfo{pages}{637--647}.
\newblock


\bibitem[\protect\citeauthoryear{Cordy and Roy}{Cordy and Roy}{2011}]%
        {NiCad:ICPC2011}
\bibfield{author}{\bibinfo{person}{James~R. Cordy} {and}
  \bibinfo{person}{Chanchal~K. Roy}.} \bibinfo{year}{2011}\natexlab{}.
\newblock \showarticletitle{The NiCad Clone Detector}. In
  \bibinfo{booktitle}{\emph{Proceedings of the 2011 IEEE 19th International
  Conference on Program Comprehension}} \emph{(\bibinfo{series}{ICPC '11})}.
  \bibinfo{publisher}{IEEE Computer Society}, \bibinfo{address}{Washington, DC,
  USA}, \bibinfo{pages}{219--220}.
\newblock


\bibitem[\protect\citeauthoryear{Duala-Ekoko and Robillard}{Duala-Ekoko and
  Robillard}{2007}]%
        {duala2007tracking}
\bibfield{author}{\bibinfo{person}{Ekwa Duala-Ekoko} {and}
  \bibinfo{person}{Martin~P Robillard}.} \bibinfo{year}{2007}\natexlab{}.
\newblock \showarticletitle{Tracking code clones in evolving software}. In
  \bibinfo{booktitle}{\emph{Software Engineering, 2007. ICSE 2007. 29th
  International Conference on}}. IEEE, \bibinfo{pages}{158--167}.
\newblock


\bibitem[\protect\citeauthoryear{Durieux, Martinez, Monperrus, Sommerard, and
  Xuan}{Durieux et~al\mbox{.}}{2015}]%
        {Durieux:CoRR2015}
\bibfield{author}{\bibinfo{person}{Thomas Durieux}, \bibinfo{person}{Matias
  Martinez}, \bibinfo{person}{Martin Monperrus}, \bibinfo{person}{Romain
  Sommerard}, {and} \bibinfo{person}{Jifeng Xuan}.}
  \bibinfo{year}{2015}\natexlab{}.
\newblock \showarticletitle{Automatic Repair of Real Bugs: An Experience Report
  on the Defects4J Dataset}.
\newblock \bibinfo{journal}{\emph{CoRR}}  \bibinfo{volume}{abs/1505.07002}
  (\bibinfo{year}{2015}).
\newblock
\urldef\tempurl%
\url{http://arxiv.org/abs/1505.07002}
\showURL{%
\tempurl}


\bibitem[\protect\citeauthoryear{Gazzola, Micucci, and Mariani}{Gazzola
  et~al\mbox{.}}{2018}]%
        {APRSurvey:2018}
\bibfield{author}{\bibinfo{person}{L. Gazzola}, \bibinfo{person}{D. Micucci},
  {and} \bibinfo{person}{L. Mariani}.} \bibinfo{year}{2018}\natexlab{}.
\newblock \showarticletitle{Automatic Software Repair: A Survey}.
\newblock \bibinfo{journal}{\emph{IEEE Transactions on Software Engineering}}
  (\bibinfo{year}{2018}), \bibinfo{pages}{1--1}.
\newblock


\bibitem[\protect\citeauthoryear{Islam, Mondal, and Roy}{Islam
  et~al\mbox{.}}{2016}]%
        {islam2016bug}
\bibfield{author}{\bibinfo{person}{Judith~F Islam},
  \bibinfo{person}{Manishankar Mondal}, {and} \bibinfo{person}{Chanchal~K
  Roy}.} \bibinfo{year}{2016}\natexlab{}.
\newblock \showarticletitle{Bug replication in code clones: An empirical
  study}. In \bibinfo{booktitle}{\emph{Software Analysis, Evolution, and
  Reengineering (SANER), 2016 IEEE 23rd International Conference on}},
  Vol.~\bibinfo{volume}{1}. IEEE, \bibinfo{pages}{68--78}.
\newblock


\bibitem[\protect\citeauthoryear{Janssen, Abreu, and van Gemund}{Janssen
  et~al\mbox{.}}{2009}]%
        {Zoltar:SINTER2009}
\bibfield{author}{\bibinfo{person}{Tom Janssen}, \bibinfo{person}{Rui Abreu},
  {and} \bibinfo{person}{Arjan~J.C. van Gemund}.}
  \bibinfo{year}{2009}\natexlab{}.
\newblock \showarticletitle{Zoltar: a spectrum-based fault localization tool}.
  In \bibinfo{booktitle}{\emph{SINTER '09: Proceedings of the 2009 ESEC/FSE
  workshop on Software integration and evolution @ runtime}}.
  \bibinfo{publisher}{ACM}, \bibinfo{address}{New York, NY, USA},
  \bibinfo{pages}{23--30}.
\newblock


\bibitem[\protect\citeauthoryear{Jiang, Xiong, Zhang, Gao, and Chen}{Jiang
  et~al\mbox{.}}{2018}]%
        {SimFix:ISSTA2018}
\bibfield{author}{\bibinfo{person}{Jiajun Jiang}, \bibinfo{person}{Yingfei
  Xiong}, \bibinfo{person}{Hongyu Zhang}, \bibinfo{person}{Qing Gao}, {and}
  \bibinfo{person}{Xiangqun Chen}.} \bibinfo{year}{2018}\natexlab{}.
\newblock \showarticletitle{Shaping Program Repair Space with Existing Patches
  and Similar Code}. In \bibinfo{booktitle}{\emph{Proceedings of the 27th ACM
  SIGSOFT International Symposium on Software Testing and Analysis}}
  \emph{(\bibinfo{series}{ISSTA 2018})}. \bibinfo{publisher}{ACM},
  \bibinfo{address}{New York, NY, USA}, \bibinfo{pages}{298--309}.
\newblock


\bibitem[\protect\citeauthoryear{Jiang, Misherghi, Su, and Glondu}{Jiang
  et~al\mbox{.}}{2007}]%
        {deckard:ICSE2007}
\bibfield{author}{\bibinfo{person}{Lingxiao Jiang}, \bibinfo{person}{Ghassan
  Misherghi}, \bibinfo{person}{Zhendong Su}, {and} \bibinfo{person}{Stephane
  Glondu}.} \bibinfo{year}{2007}\natexlab{}.
\newblock \showarticletitle{Deckard: Scalable and accurate tree-based detection
  of code clones}. In \bibinfo{booktitle}{\emph{Proceedings of the 29th
  international conference on Software Engineering}}. IEEE Computer Society,
  \bibinfo{pages}{96--105}.
\newblock


\bibitem[\protect\citeauthoryear{Jones, Harrold, and Stasko}{Jones
  et~al\mbox{.}}{2002}]%
        {Tarantula:ICSE2002}
\bibfield{author}{\bibinfo{person}{James~A. Jones}, \bibinfo{person}{Mary~Jean
  Harrold}, {and} \bibinfo{person}{John Stasko}.}
  \bibinfo{year}{2002}\natexlab{}.
\newblock \showarticletitle{Visualization of Test Information to Assist Fault
  Localization}. In \bibinfo{booktitle}{\emph{Proceedings of the 24th
  International Conference on Software Engineering}}
  \emph{(\bibinfo{series}{ICSE '02})}. \bibinfo{publisher}{ACM},
  \bibinfo{address}{New York, NY, USA}, \bibinfo{pages}{467--477}.
\newblock


\bibitem[\protect\citeauthoryear{Juergens, Deissenboeck, Hummel, and
  Wagner}{Juergens et~al\mbox{.}}{2009}]%
        {juergens2009code}
\bibfield{author}{\bibinfo{person}{Elmar Juergens}, \bibinfo{person}{Florian
  Deissenboeck}, \bibinfo{person}{Benjamin Hummel}, {and}
  \bibinfo{person}{Stefan Wagner}.} \bibinfo{year}{2009}\natexlab{}.
\newblock \showarticletitle{Do code clones matter?}. In
  \bibinfo{booktitle}{\emph{Software Engineering, 2009. ICSE 2009. IEEE 31st
  International Conference on}}. IEEE, \bibinfo{pages}{485--495}.
\newblock


\bibitem[\protect\citeauthoryear{Just, Jalali, and Ernst}{Just
  et~al\mbox{.}}{2014}]%
        {just2014defects4j}
\bibfield{author}{\bibinfo{person}{Ren{\'e} Just}, \bibinfo{person}{Darioush
  Jalali}, {and} \bibinfo{person}{Michael~D Ernst}.}
  \bibinfo{year}{2014}\natexlab{}.
\newblock \showarticletitle{Defects4J: A database of existing faults to enable
  controlled testing studies for Java programs}. In
  \bibinfo{booktitle}{\emph{Proceedings of the 2014 International Symposium on
  Software Testing and Analysis}}. ACM, \bibinfo{pages}{437--440}.
\newblock


\bibitem[\protect\citeauthoryear{Kamiya, Kusumoto, and Inoue}{Kamiya
  et~al\mbox{.}}{2002}]%
        {CCFinder:TSE2002}
\bibfield{author}{\bibinfo{person}{Toshihiro Kamiya}, \bibinfo{person}{Shinji
  Kusumoto}, {and} \bibinfo{person}{Katsuro Inoue}.}
  \bibinfo{year}{2002}\natexlab{}.
\newblock \showarticletitle{CCFinder: A Multilinguistic Token-based Code Clone
  Detection System for Large Scale Source Code}.
\newblock \bibinfo{journal}{\emph{IEEE Trans. Softw. Eng.}}
  \bibinfo{volume}{28}, \bibinfo{number}{7} (\bibinfo{date}{July}
  \bibinfo{year}{2002}), \bibinfo{pages}{654--670}.
\newblock
\showISSN{0098-5589}


\bibitem[\protect\citeauthoryear{Ke, Stolee, {Le Goues}, and Brun}{Ke
  et~al\mbox{.}}{2015}]%
        {SearchRepair:ASE2015}
\bibfield{author}{\bibinfo{person}{Y. Ke}, \bibinfo{person}{K.~T. Stolee},
  \bibinfo{person}{C. {Le Goues}}, {and} \bibinfo{person}{Y. Brun}.}
  \bibinfo{year}{2015}\natexlab{}.
\newblock \showarticletitle{Repairing Programs with Semantic Code Search}. In
  \bibinfo{booktitle}{\emph{2015 30th IEEE/ACM International Conference on
  Automated Software Engineering (ASE)}}. \bibinfo{pages}{295--306}.
\newblock


\bibitem[\protect\citeauthoryear{Kim, Nam, Song, and Kim}{Kim
  et~al\mbox{.}}{2013}]%
        {PAR:ICSE2013}
\bibfield{author}{\bibinfo{person}{Dongsun Kim}, \bibinfo{person}{Jaechang
  Nam}, \bibinfo{person}{Jaewoo Song}, {and} \bibinfo{person}{Sunghun Kim}.}
  \bibinfo{year}{2013}\natexlab{}.
\newblock \showarticletitle{Automatic Patch Generation Learned from
  Human-written Patches}. In \bibinfo{booktitle}{\emph{Proceedings of the 2013
  International Conference on Software Engineering}}
  \emph{(\bibinfo{series}{ICSE '13})}. \bibinfo{publisher}{IEEE Press},
  \bibinfo{address}{Piscataway, NJ, USA}, \bibinfo{pages}{802--811}.
\newblock


\bibitem[\protect\citeauthoryear{Kim and Notkin}{Kim and Notkin}{2005}]%
        {kim2005using}
\bibfield{author}{\bibinfo{person}{Miryung Kim} {and} \bibinfo{person}{David
  Notkin}.} \bibinfo{year}{2005}\natexlab{}.
\newblock \showarticletitle{Using a clone genealogy extractor for understanding
  and supporting evolution of code clones}. In \bibinfo{booktitle}{\emph{ACM
  SIGSOFT Software Engineering Notes}}, Vol.~\bibinfo{volume}{30}. ACM,
  \bibinfo{pages}{1--5}.
\newblock


\bibitem[\protect\citeauthoryear{Kim and Notkin}{Kim and Notkin}{2009}]%
        {Kim:ICSE2009}
\bibfield{author}{\bibinfo{person}{Miryung Kim} {and} \bibinfo{person}{David
  Notkin}.} \bibinfo{year}{2009}\natexlab{}.
\newblock \showarticletitle{Discovering and Representing Systematic Code
  Changes}. In \bibinfo{booktitle}{\emph{Proceedings of the 31st International
  Conference on Software Engineering}} \emph{(\bibinfo{series}{ICSE '09})}.
  \bibinfo{publisher}{IEEE Computer Society}, \bibinfo{address}{Washington, DC,
  USA}, \bibinfo{pages}{309--319}.
\newblock


\bibitem[\protect\citeauthoryear{Le, Lo, and {Le Goues}}{Le
  et~al\mbox{.}}{2016}]%
        {HDRepair:SANER2016}
\bibfield{author}{\bibinfo{person}{X.~B.~D. Le}, \bibinfo{person}{D. Lo}, {and}
  \bibinfo{person}{C. {Le Goues}}.} \bibinfo{year}{2016}\natexlab{}.
\newblock \showarticletitle{{History Driven Program Repair}}. In
  \bibinfo{booktitle}{\emph{2016 IEEE 23rd International Conference on Software
  Analysis, Evolution, and Reengineering (SANER)}}, Vol.~\bibinfo{volume}{1}.
  \bibinfo{publisher}{IEEE Press}, \bibinfo{address}{Piscataway, NJ, USA},
  \bibinfo{pages}{213--224}.
\newblock


\bibitem[\protect\citeauthoryear{{Le Goues}, Dewey-Vogt, Forrest, and
  Weimer}{{Le Goues} et~al\mbox{.}}{2012}]%
        {GenProg:ICSE2012}
\bibfield{author}{\bibinfo{person}{Claire {Le Goues}}, \bibinfo{person}{Michael
  Dewey-Vogt}, \bibinfo{person}{Stephanie Forrest}, {and}
  \bibinfo{person}{Westley Weimer}.} \bibinfo{year}{2012}\natexlab{}.
\newblock \showarticletitle{A Systematic Study of Automated Program Repair:
  Fixing 55 out of 105 Bugs for \$8 Each}. In
  \bibinfo{booktitle}{\emph{Proceedings of the 34th International Conference on
  Software Engineering}} \emph{(\bibinfo{series}{ICSE '12})}.
  \bibinfo{publisher}{IEEE Press}, \bibinfo{address}{Piscataway, NJ, USA},
  \bibinfo{pages}{3--13}.
\newblock


\bibitem[\protect\citeauthoryear{Le~Goues, Forrest, and Weimer}{Le~Goues
  et~al\mbox{.}}{2013}]%
        {le2013current}
\bibfield{author}{\bibinfo{person}{Claire Le~Goues}, \bibinfo{person}{Stephanie
  Forrest}, {and} \bibinfo{person}{Westley Weimer}.}
  \bibinfo{year}{2013}\natexlab{}.
\newblock \showarticletitle{Current challenges in automatic software repair}.
\newblock \bibinfo{journal}{\emph{Software quality journal}}
  \bibinfo{volume}{21}, \bibinfo{number}{3} (\bibinfo{year}{2013}),
  \bibinfo{pages}{421--443}.
\newblock


\bibitem[\protect\citeauthoryear{Long, Amidon, and Rinard}{Long
  et~al\mbox{.}}{2017}]%
        {Genesis:FSE2017}
\bibfield{author}{\bibinfo{person}{Fan Long}, \bibinfo{person}{Peter Amidon},
  {and} \bibinfo{person}{Martin Rinard}.} \bibinfo{year}{2017}\natexlab{}.
\newblock \showarticletitle{Automatic Inference of Code Transforms for Patch
  Generation}. In \bibinfo{booktitle}{\emph{Proceedings of the 2017 11th Joint
  Meeting on Foundations of Software Engineering}}
  \emph{(\bibinfo{series}{ESEC/FSE 2017})}. \bibinfo{publisher}{ACM},
  \bibinfo{address}{New York, NY, USA}, \bibinfo{pages}{727--739}.
\newblock


\bibitem[\protect\citeauthoryear{Long and Rinard}{Long and Rinard}{2015}]%
        {SPR:FSE2015}
\bibfield{author}{\bibinfo{person}{Fan Long} {and} \bibinfo{person}{Martin
  Rinard}.} \bibinfo{year}{2015}\natexlab{}.
\newblock \showarticletitle{Staged Program Repair with Condition Synthesis}. In
  \bibinfo{booktitle}{\emph{Proceedings of the 2015 10th Joint Meeting on
  Foundations of Software Engineering}} \emph{(\bibinfo{series}{ESEC/FSE
  2015})}. \bibinfo{publisher}{ACM}, \bibinfo{address}{New York, NY, USA},
  \bibinfo{pages}{166--178}.
\newblock


\bibitem[\protect\citeauthoryear{Long and Rinard}{Long and Rinard}{2016}]%
        {Prophet:POPL2016}
\bibfield{author}{\bibinfo{person}{Fan Long} {and} \bibinfo{person}{Martin
  Rinard}.} \bibinfo{year}{2016}\natexlab{}.
\newblock \showarticletitle{Automatic Patch Generation by Learning Correct
  Code}. In \bibinfo{booktitle}{\emph{Proceedings of the 43rd Annual ACM
  SIGPLAN-SIGACT Symposium on Principles of Programming Languages}}
  \emph{(\bibinfo{series}{POPL '16})}. \bibinfo{publisher}{ACM},
  \bibinfo{address}{New York, NY, USA}, \bibinfo{pages}{298--312}.
\newblock


\bibitem[\protect\citeauthoryear{Mechtaev, Yi, and Roychoudhury}{Mechtaev
  et~al\mbox{.}}{2016}]%
        {Angelix:ICSE2016}
\bibfield{author}{\bibinfo{person}{Sergey Mechtaev}, \bibinfo{person}{Jooyong
  Yi}, {and} \bibinfo{person}{Abhik Roychoudhury}.}
  \bibinfo{year}{2016}\natexlab{}.
\newblock \showarticletitle{{Angelix: Scalable Multiline Program Patch
  Synthesis via Symbolic Analysis}}. In \bibinfo{booktitle}{\emph{Proceedings
  of the 38th International Conference on Software Engineering}}
  \emph{(\bibinfo{series}{ICSE '16})}. \bibinfo{publisher}{ACM},
  \bibinfo{address}{New York, NY, USA}, \bibinfo{pages}{691--701}.
\newblock


\bibitem[\protect\citeauthoryear{Meng, Hua, Kim, and McKinley}{Meng
  et~al\mbox{.}}{2015}]%
        {Rase:ICSE2015}
\bibfield{author}{\bibinfo{person}{Na Meng}, \bibinfo{person}{Lisa Hua},
  \bibinfo{person}{Miryung Kim}, {and} \bibinfo{person}{Kathryn~S. McKinley}.}
  \bibinfo{year}{2015}\natexlab{}.
\newblock \showarticletitle{Does Automated Refactoring Obviate Systematic
  Editing?}. In \bibinfo{booktitle}{\emph{Proceedings of the 37th International
  Conference on Software Engineering - Volume 1}} \emph{(\bibinfo{series}{ICSE
  '15})}. \bibinfo{publisher}{IEEE Press}, \bibinfo{address}{Piscataway, NJ,
  USA}, \bibinfo{pages}{392--402}.
\newblock


\bibitem[\protect\citeauthoryear{Meng, Kim, and McKinley}{Meng
  et~al\mbox{.}}{2011}]%
        {Meng:PLDI2011}
\bibfield{author}{\bibinfo{person}{Na Meng}, \bibinfo{person}{Miryung Kim},
  {and} \bibinfo{person}{Kathryn~S. McKinley}.}
  \bibinfo{year}{2011}\natexlab{}.
\newblock \showarticletitle{Systematic Editing: Generating Program
  Transformations from an Example}. In \bibinfo{booktitle}{\emph{Proceedings of
  the 32Nd ACM SIGPLAN Conference on Programming Language Design and
  Implementation}} \emph{(\bibinfo{series}{PLDI '11})}.
  \bibinfo{publisher}{ACM}, \bibinfo{address}{New York, NY, USA},
  \bibinfo{pages}{329--342}.
\newblock


\bibitem[\protect\citeauthoryear{Meng, Kim, and McKinley}{Meng
  et~al\mbox{.}}{2013}]%
        {LASE:ICSE2013}
\bibfield{author}{\bibinfo{person}{Na Meng}, \bibinfo{person}{Miryung Kim},
  {and} \bibinfo{person}{Kathryn~S. McKinley}.}
  \bibinfo{year}{2013}\natexlab{}.
\newblock \showarticletitle{LASE: Locating and Applying Systematic Edits by
  Learning from Examples}. In \bibinfo{booktitle}{\emph{Proceedings of the 2013
  International Conference on Software Engineering}}
  \emph{(\bibinfo{series}{ICSE '13})}. \bibinfo{publisher}{IEEE Press},
  \bibinfo{address}{Piscataway, NJ, USA}, \bibinfo{pages}{502--511}.
\newblock


\bibitem[\protect\citeauthoryear{Monperrus}{Monperrus}{2018}]%
        {Monperrus:2018}
\bibfield{author}{\bibinfo{person}{Martin Monperrus}.}
  \bibinfo{year}{2018}\natexlab{}.
\newblock \showarticletitle{Automatic Software Repair: A Bibliography}.
\newblock \bibinfo{journal}{\emph{Comput. Surveys}} \bibinfo{volume}{51},
  \bibinfo{number}{1}, Article \bibinfo{articleno}{17} (\bibinfo{date}{Jan.}
  \bibinfo{year}{2018}), \bibinfo{numpages}{24}~pages.
\newblock


\bibitem[\protect\citeauthoryear{Qi, Mao, Lei, Dai, and Wang}{Qi
  et~al\mbox{.}}{2014}]%
        {RSRepair:ICSE2014}
\bibfield{author}{\bibinfo{person}{Yuhua Qi}, \bibinfo{person}{Xiaoguang Mao},
  \bibinfo{person}{Yan Lei}, \bibinfo{person}{Ziying Dai}, {and}
  \bibinfo{person}{Chengsong Wang}.} \bibinfo{year}{2014}\natexlab{}.
\newblock \showarticletitle{The Strength of Random Search on Automated Program
  Repair}. In \bibinfo{booktitle}{\emph{Proceedings of the 36th International
  Conference on Software Engineering}} \emph{(\bibinfo{series}{ICSE 2014})}.
  \bibinfo{publisher}{ACM}, \bibinfo{address}{New York, NY, USA},
  \bibinfo{pages}{254--265}.
\newblock


\bibitem[\protect\citeauthoryear{Qi, Long, Achour, and Rinard}{Qi
  et~al\mbox{.}}{2015}]%
        {Kali:ISSTA2015}
\bibfield{author}{\bibinfo{person}{Zichao Qi}, \bibinfo{person}{Fan Long},
  \bibinfo{person}{Sara Achour}, {and} \bibinfo{person}{Martin Rinard}.}
  \bibinfo{year}{2015}\natexlab{}.
\newblock \showarticletitle{An Analysis of Patch Plausibility and Correctness
  for Generate-and-validate Patch Generation Systems}. In
  \bibinfo{booktitle}{\emph{Proceedings of the 2015 International Symposium on
  Software Testing and Analysis}} \emph{(\bibinfo{series}{ISSTA 2015})}.
  \bibinfo{publisher}{ACM}, \bibinfo{address}{New York, NY, USA},
  \bibinfo{pages}{24--36}.
\newblock


\bibitem[\protect\citeauthoryear{Rieger, Ducasse, and Lanza}{Rieger
  et~al\mbox{.}}{2004}]%
        {rieger2004insights}
\bibfield{author}{\bibinfo{person}{Matthias Rieger},
  \bibinfo{person}{St{\'e}phane Ducasse}, {and} \bibinfo{person}{Michele
  Lanza}.} \bibinfo{year}{2004}\natexlab{}.
\newblock \showarticletitle{Insights into system-wide code duplication}. In
  \bibinfo{booktitle}{\emph{Reverse Engineering, 2004. Proceedings. 11th
  Working Conference on}}. IEEE, \bibinfo{pages}{100--109}.
\newblock


\bibitem[\protect\citeauthoryear{Roy and Cordy}{Roy and Cordy}{2007}]%
        {roy2007survey}
\bibfield{author}{\bibinfo{person}{Chanchal~Kumar Roy} {and}
  \bibinfo{person}{James~R Cordy}.} \bibinfo{year}{2007}\natexlab{}.
\newblock \showarticletitle{A survey on software clone detection research}.
\newblock \bibinfo{journal}{\emph{Queen?s School of Computing TR}}
  \bibinfo{volume}{541}, \bibinfo{number}{115} (\bibinfo{year}{2007}),
  \bibinfo{pages}{64--68}.
\newblock


\bibitem[\protect\citeauthoryear{Roy and Cordy}{Roy and Cordy}{2008}]%
        {roy2008empirical}
\bibfield{author}{\bibinfo{person}{Chanchal~K Roy} {and}
  \bibinfo{person}{James~R Cordy}.} \bibinfo{year}{2008}\natexlab{}.
\newblock \showarticletitle{An empirical study of function clones in open
  source software}. In \bibinfo{booktitle}{\emph{2008 15th Working Conference
  on Reverse Engineering}}. IEEE, \bibinfo{pages}{81--90}.
\newblock


\bibitem[\protect\citeauthoryear{Saha, Lyu, Lam, Yoshida, and Prasad}{Saha
  et~al\mbox{.}}{2018}]%
        {saha2018bugs}
\bibfield{author}{\bibinfo{person}{Ripon~K Saha}, \bibinfo{person}{Yingjun
  Lyu}, \bibinfo{person}{Wing Lam}, \bibinfo{person}{Hiroaki Yoshida}, {and}
  \bibinfo{person}{Mukul~R Prasad}.} \bibinfo{year}{2018}\natexlab{}.
\newblock \showarticletitle{Bugs. jar: a large-scale, diverse dataset of
  real-world Java bugs}. In \bibinfo{booktitle}{\emph{Proceedings of the 15th
  International Conference on Mining Software Repositories}}. ACM,
  \bibinfo{pages}{10--13}.
\newblock


\bibitem[\protect\citeauthoryear{Saha, Lyu, Yoshida, and Prasad}{Saha
  et~al\mbox{.}}{2017}]%
        {Elixir:ASE2017}
\bibfield{author}{\bibinfo{person}{Ripon~K. Saha}, \bibinfo{person}{Yingjun
  Lyu}, \bibinfo{person}{Hiroaki Yoshida}, {and} \bibinfo{person}{Mukul~R.
  Prasad}.} \bibinfo{year}{2017}\natexlab{}.
\newblock \showarticletitle{ELIXIR: Effective Object Oriented Program Repair}.
  In \bibinfo{booktitle}{\emph{Proceedings of the 32Nd IEEE/ACM International
  Conference on Automated Software Engineering}} \emph{(\bibinfo{series}{ASE
  2017})}. \bibinfo{publisher}{IEEE Press}, \bibinfo{address}{Piscataway, NJ,
  USA}, \bibinfo{pages}{648--659}.
\newblock


\bibitem[\protect\citeauthoryear{Saha, Saha, and Prasad}{Saha
  et~al\mbox{.}}{2019}]%
        {Hercules:ICSE2019}
\bibfield{author}{\bibinfo{person}{Seemanta Saha}, \bibinfo{person}{Ripon~K.
  Saha}, {and} \bibinfo{person}{Mukul~R. Prasad}.}
  \bibinfo{year}{2019}\natexlab{}.
\newblock \showarticletitle{Harnessing Evolution for Multi-Hunk Program
  Repair}. In \bibinfo{booktitle}{\emph{Proceedings of the IEEE/ACM 41st
  International Conference on Software Engineering}}
  \emph{(\bibinfo{series}{ICSE '19})}. \bibinfo{publisher}{IEEE Press},
  \bibinfo{address}{Piscataway, NJ, USA}, \bibinfo{pages}{13--24}.
\newblock


\bibitem[\protect\citeauthoryear{Sidiroglou-Douskos, Lahtinen, Long, and
  Rinard}{Sidiroglou-Douskos et~al\mbox{.}}{2015}]%
        {CodePhage:PLDI2015}
\bibfield{author}{\bibinfo{person}{Stelios Sidiroglou-Douskos},
  \bibinfo{person}{Eric Lahtinen}, \bibinfo{person}{Fan Long}, {and}
  \bibinfo{person}{Martin Rinard}.} \bibinfo{year}{2015}\natexlab{}.
\newblock \showarticletitle{Automatic Error Elimination by Horizontal Code
  Transfer Across Multiple Applications}. In
  \bibinfo{booktitle}{\emph{Proceedings of the 36th ACM SIGPLAN Conference on
  Programming Language Design and Implementation}} \emph{(\bibinfo{series}{PLDI
  '15})}. \bibinfo{publisher}{ACM}, \bibinfo{address}{New York, NY, USA},
  \bibinfo{pages}{43--54}.
\newblock


\bibitem[\protect\citeauthoryear{Tan and Roychoudhury}{Tan and
  Roychoudhury}{2015}]%
        {Relifix:ICSE2015}
\bibfield{author}{\bibinfo{person}{Shin~Hwei Tan} {and} \bibinfo{person}{Abhik
  Roychoudhury}.} \bibinfo{year}{2015}\natexlab{}.
\newblock \showarticletitle{Relifix: Automated Repair of Software Regressions}.
  In \bibinfo{booktitle}{\emph{Proceedings of the 37th International Conference
  on Software Engineering - Volume 1}} \emph{(\bibinfo{series}{ICSE '15})}.
  \bibinfo{publisher}{IEEE Press}, \bibinfo{address}{Piscataway, NJ, USA},
  \bibinfo{pages}{471--482}.
\newblock


\bibitem[\protect\citeauthoryear{Tan, Yoshida, Prasad, and Roychoudhury}{Tan
  et~al\mbox{.}}{2016}]%
        {antiPatterns:FSE2016}
\bibfield{author}{\bibinfo{person}{Shin~Hwei Tan}, \bibinfo{person}{Hiroaki
  Yoshida}, \bibinfo{person}{Mukul~R. Prasad}, {and} \bibinfo{person}{Abhik
  Roychoudhury}.} \bibinfo{year}{2016}\natexlab{}.
\newblock \showarticletitle{Anti-patterns in Search-based Program Repair}. In
  \bibinfo{booktitle}{\emph{Proceedings of the 2016 24th ACM SIGSOFT
  International Symposium on Foundations of Software Engineering}}
  \emph{(\bibinfo{series}{FSE 2016})}. \bibinfo{publisher}{ACM},
  \bibinfo{address}{New York, NY, USA}, \bibinfo{pages}{727--738}.
\newblock


\bibitem[\protect\citeauthoryear{Weimer, Fry, and Forrest}{Weimer
  et~al\mbox{.}}{2013}]%
        {AE:ASE2013}
\bibfield{author}{\bibinfo{person}{W. Weimer}, \bibinfo{person}{Z.~P. Fry},
  {and} \bibinfo{person}{S. Forrest}.} \bibinfo{year}{2013}\natexlab{}.
\newblock \showarticletitle{Leveraging program equivalence for adaptive program
  repair: Models and first results}. In \bibinfo{booktitle}{\emph{Automated
  Software Engineering (ASE), 2013 IEEE/ACM 28th International Conference on}}.
  \bibinfo{publisher}{IEEE Press}, \bibinfo{address}{Piscataway, NJ, USA},
  \bibinfo{pages}{356--366}.
\newblock


\bibitem[\protect\citeauthoryear{Wen, Chen, Wu, Hao, and Cheung}{Wen
  et~al\mbox{.}}{2018}]%
        {CapGen:ICSE2018}
\bibfield{author}{\bibinfo{person}{Ming Wen}, \bibinfo{person}{Junjie Chen},
  \bibinfo{person}{Rongxin Wu}, \bibinfo{person}{Dan Hao}, {and}
  \bibinfo{person}{Shing-Chi Cheung}.} \bibinfo{year}{2018}\natexlab{}.
\newblock \showarticletitle{Context-aware Patch Generation for Better Automated
  Program Repair}. In \bibinfo{booktitle}{\emph{Proceedings of the 40th
  International Conference on Software Engineering}}
  \emph{(\bibinfo{series}{ICSE '18})}. \bibinfo{publisher}{ACM},
  \bibinfo{address}{New York, NY, USA}, \bibinfo{pages}{1--11}.
\newblock


\bibitem[\protect\citeauthoryear{Xin and Reiss}{Xin and Reiss}{2017}]%
        {ssFix:ASE2017}
\bibfield{author}{\bibinfo{person}{Qi Xin} {and} \bibinfo{person}{Steven~P.
  Reiss}.} \bibinfo{year}{2017}\natexlab{}.
\newblock \showarticletitle{Leveraging Syntax-related Code for Automated
  Program Repair}. In \bibinfo{booktitle}{\emph{Proceedings of the 32Nd
  IEEE/ACM International Conference on Automated Software Engineering}}
  \emph{(\bibinfo{series}{ASE 2017})}. \bibinfo{publisher}{IEEE Press},
  \bibinfo{address}{Piscataway, NJ, USA}, \bibinfo{pages}{660--670}.
\newblock


\bibitem[\protect\citeauthoryear{Xiong, Wang, Yan, Zhang, Han, Huang, and
  Zhang}{Xiong et~al\mbox{.}}{2017}]%
        {ACS:ICSE2017}
\bibfield{author}{\bibinfo{person}{Yingfei Xiong}, \bibinfo{person}{Jie Wang},
  \bibinfo{person}{Runfa Yan}, \bibinfo{person}{Jiachen Zhang},
  \bibinfo{person}{Shi Han}, \bibinfo{person}{Gang Huang}, {and}
  \bibinfo{person}{Lu Zhang}.} \bibinfo{year}{2017}\natexlab{}.
\newblock \showarticletitle{Precise Condition Synthesis for Program Repair}. In
  \bibinfo{booktitle}{\emph{Proceedings of the 39th International Conference on
  Software Engineering}} \emph{(\bibinfo{series}{ICSE '17})}.
  \bibinfo{publisher}{IEEE Press}, \bibinfo{address}{Piscataway, NJ, USA},
  \bibinfo{pages}{416--426}.
\newblock


\bibitem[\protect\citeauthoryear{Zhang and Shasha}{Zhang and Shasha}{1989}]%
        {zhang1989simple}
\bibfield{author}{\bibinfo{person}{Kaizhong Zhang} {and}
  \bibinfo{person}{Dennis Shasha}.} \bibinfo{year}{1989}\natexlab{}.
\newblock \showarticletitle{Simple fast algorithms for the editing distance
  between trees and related problems}.
\newblock \bibinfo{journal}{\emph{SIAM journal on computing}}
  \bibinfo{volume}{18}, \bibinfo{number}{6} (\bibinfo{year}{1989}),
  \bibinfo{pages}{1245--1262}.
\newblock


\bibitem[\protect\citeauthoryear{Zhong and Su}{Zhong and Su}{2015}]%
        {realbug:ICSE2015}
\bibfield{author}{\bibinfo{person}{Hao Zhong} {and} \bibinfo{person}{Zhendong
  Su}.} \bibinfo{year}{2015}\natexlab{}.
\newblock \showarticletitle{An empirical study on real bug fixes}. In
  \bibinfo{booktitle}{\emph{Proceedings of the 37th International Conference on
  Software Engineering-Volume 1}}. IEEE Press, \bibinfo{pages}{913--923}.
\newblock


\end{thebibliography}

\end{document}